\documentclass[a4paper,fleqn]{cas-sc}

\usepackage[authoryear]{natbib}
\usepackage{graphicx}
\usepackage{subcaption}
\usepackage{float} 
\usepackage{placeins}
\usepackage{capt-of}
\usepackage{url}
\usepackage{lineno}

\def\tsc#1{\csdef{#1}{\textsc{\lowercase{#1}}\xspace}}
\tsc{WGM}
\tsc{QE}
\newif\ifanonym
\anonymfalse
\begin{document}
%\linenumbers
\let\WriteBookmarks\relax
\def\floatpagepagefraction{1}
\def\textpagefraction{.001}

\ifanonym
    \shorttitle{Machine Learning Meteor Classification}
    \shortauthors{Anonymous et al.} 
\else
    \shorttitle{Machine Learning Meteor Classification}
    \shortauthors{Hemmelgarn et al.} 
\fi

\title[mode=title]{A machine learning approach to meteor classification}  

\ifanonym
    \author{Anonymous Author(s)}
    \affiliation{organization={Anonymous Institution}}
\else
    \tnotetext[1]{Corresponding Author} 
    \author[1, 2]{Samantha Hemmelgarn}[orcid=0009-0006-1160-3829]
    \cormark[1]
    \ead{shemmelgarn@lowell.edu}
    \credit{Conceptualization; Data curation; Formal analysis; Investigation; Methodology; Project administration; Software; Validation; Visualization; Writing - original draft; and Writing - review \& editing}

    \author[2]{Nicholas Moskovitz}[orcid=0000-0001-6765-6336]
    \ead{nmosko@lowell.edu}
    \credit{Conceptualization; Data curation; Funding acquisition; Investigation; Methodology; Project administration; Resources; Supervision; Validation; and Writing - review \& editing}
    
    \author[3, 4]{Denis Vida}[orcid=0000-0003-4166-8704]
    \ead{dvida@uwo.ca}
    \credit{Conceptualization; Data curation; Funding acquisition; Investigation; Methodology; Project administration; Resources; Supervision; Validation; and Writing - review \& editing}

    % Address/affiliations
    \affiliation[1]{organization={Department of Astronomy and Planetary Science, Northern Arizona University},
        addressline={527 S Beaver St, Flagstaff}, 
        city={Flagstaff}, 
        state={AZ},
        postcode={86011},
        country={USA}}
    \affiliation[2]{organization={Lowell Observatory},
        addressline={1400 West Mars Hill Road}, 
        city={Flagstaff}, 
        state={AZ},
        postcode={86001},
        country={USA}}
    \affiliation[3]{organization={Department of Physics and Astronomy, The University of Western Ontario}, 
            city={London, Ontario},
            country={Canada}}
    \affiliation[4]{organization={Institute for Earth and Space Exploration, The University of Western Ontario}, 
            city={London, Ontario},
            country={Canada}}
\fi

\begin{abstract}
We use machine learning to develop a framework for classifying meteoroids based on 13 directly observed parameters from the Global Meteor Network. This method adds depth to the $K_{b}$ parameter, which uses only three parameters. We employ a semi-qualitative approach using 28,177 meteor events observed in 2023 by the Lowell Observatory Cameras for All-Sky Meteor Surveillance (LO-CAMS) network to evaluate multiple normalization, dimensionality-reduction, and clustering algorithms. We find that a combination of Factor Analysis (FA) and a Gaussian Mixture Model (GMM) results in clusters most consistent with traditional models. Three FA-derived factors corresponding to meteoroid kinematics, activation thresholds, and size/geometry effects describe the underlying structure of meteoroid behavior. The activation factor emerged as the most discriminating factor distinguishing whether a meteor is of asteroidal or cometary origin. Resulting 3, 6, and 11 cluster models reveal progressively finer compositional structure, from broad physical regimes to detailed subdivisions within cometary and asteroidal populations. From these results, we introduce a physically motivated hardness classification scheme: $H_{\mathrm{class}}$. $H_{\mathrm{class}}$ is a data-driven extension of $K_{b}$ which physically interprets clusters in terms of the densest iron meteoroids down to the softest cometary material. Application to nine well-studied meteor showers and analysis of clusters in orbital space aids in the physical interpretation of $H_{\mathrm{class}}$ groups. The $H_{\mathrm{class}}$ model is supported by an analytical FA–GMM formulation that enables application to future datasets. Our results demonstrate that machine learning methods can extract compositional information from modern optical meteor datasets at scale and offers a new framework for interpreting meteoroid populations.
\end{abstract}

% Keywords

\maketitle

% Main text
\section{Introduction} \label{sec:intro}

Meteoroids are natural interplanetary particles which ablate when they enter a gaseous atmosphere, producing light and ionization among other things \citep{borovicka2015}. Meteoroids typically come from one of two types of bodies: comets or asteroids; thus giving them distinct properties that are indicative of their parent bodies. Cometary meteoroids are considered to be desiccated remnants of low-density cometary material, essentially thermally processed fractal dust aggregates \citep{Fulle_2017}. Heat from the sun causes volatiles to sublimate from the surfaces of comets, ejecting dust from the nucleus, and creating a stream of particles \citep{Vaubaillon_2005-I, EGAL2019123}. Understanding this process is crucial for determining the origins of dust in the solar system. For example, meteoroid streams from Long-Period Comets (LPCs) can be traced back to their sources in the outer Solar System to guide searches for their parent bodies \citep{Hemmelgarn2024}. On the other hand, asteroidal meteoroid streams are believed to be formed through asteroid collisions \citep{jenniskens2006} or thermal cracking of the asteroid surface \citep{Jewitt_2010, MacLennan_2024}, although the exact mechanism of ejection is an active area of research \citep{Ryabova_2015, Jo_2024}.

Observing and analyzing the light emitted from a meteoroid as it ablates in the Earth's atmosphere informs its composition and evolutionary history. Spectroscopic analysis of meteors allows compositional classification, however it requires complex data collection, reduction, and modeling \citep{borovivcka2005survey, vojavcek2015catalogue}. Advanced meteoroid fragmentation models (e.g \cite{Borovicka_2007, Kikwaya_2011, Buccongello_2024, Vida_2024, Vovk_2025}) can provide insights on the physical parameters of meteors that do not survive their journey through the atmosphere. These models generally work by fitting a complex numerical ablation model to meteor light curve and deceleration data. This approach is computationally intensive, requires accurate manual measurements, and has thus far only been applied to datasets of a few dozen meteors at a time.  

With the development and wide-spread availability of technologies such as CCD and CMOS sensors, automated camera networks began in the early 2000s to observe the sky on a nightly basis for meteors. Networks such as Cameras for All-Sky Meteor Surveillance \citep[CAMS;][]{cams}, the Global Meteor Network \citep[GMN;][]{gmn, vida_2021, vida_2022}, the Desert Fireball Network \citep[DFN;][]{DEVILLEPOIX2020105036}, the International Meteor Organization Network \citep[IMO;][]{IMO_2001, IMO_2014}, and the Fireball Recovery and InterPlanetary Obsevation Network \citep[FRIPON;][]{Colas_2021} have curated the orbits of $\sim$3 million meteors \citep{cams, gmn} since routine observations began in the early 2010's. As automated meteor camera networks continue to grow, more scalable classification and characterization methods that rely only on what we directly observe are needed.

Classifying meteoroids based on observed physical properties is crucial for constraining their evolutionary history and establishing links to parent bodies. Current classification methods typically rely on simplified physical models, such as single-body ablation theory, or utilize a limited set of observable characteristics. A standard metric for smaller meteors is the $K_{b}$ parameter \citep{Ceplecha_1958, ceplecha1988}, which estimates material strength based on the meteoroid's ability to penetrate the atmosphere. The parameter is defined as:

\begin{equation}
    \label{k_b_87}
        K_b = \log \rho_\mathrm{beg} + \frac{5}{2}\log V_\mathrm{init} - \frac{1}{2} \log \cos \theta_z
\end{equation}

Where $\rho_\mathrm{beg}$ is the atmospheric density at the beginning of the trajectory (in $\mathrm{g\ cm^{-3}}$), $V_\mathrm{init}$ is the initial velocity (in $\mathrm{cm\ s^{-1}}$), and $\theta_z$ is the zenith angle. Empirically, $K_{b}$ defines categories of meteoroid populations by strength: values $\geq 8.0$ indicate asteroidal material, $7.3 \leq K_{b} < 8.0$ suggest carbonaceous origins, and $K_{b} < 7.3$ point to cometary sources. Harder, and therefore denser, asteroidal material refers to ordinary chondrites with mean bulk densities of 3700 $\mathrm{kg\ m^{-3}}$, carbonaceous objects average 2000 $\mathrm{kg\ m^{-3}}$, and the softest cometary meteors are closer to 1000 $\mathrm{kg\ m^{-3}}$ \citep{ceplecha1988}. However, the $K_{b}$ criterion is limited by its reliance on broad assumptions regarding meteoroid structure and a small number of entry observables. This restricts the physical fidelity of the method, particularly when analyzing the diverse datasets produced by modern automated networks. 

Alternative strength metrics, such as $PE$ \citep{Ceplecha_and_McCrosky_1976} and $P_f$ \citep{Borovicka_2022}, utilize fireball end height or maximum dynamic pressure as proxies for strength, normalizing for velocity, mass, and entry angle. While effective for decimeter-sized fireballs, these methods are ill-suited for the millimeter-to-centimeter scale meteoroids that dominate the data stream from low-light camera networks. At this smaller scale, mass loss is driven by the continuous erosion of surface grains rather than the discrete fragmentation events typical of larger bodies \citep{Borovicka_2007}. Despite the structural assumptions inherent in $K_{b}$, it remains the preferred metric for smaller meteors as it relies on properties observable at the onset of the event.

Machine learning (ML) algorithms present the opportunity to apply modern techniques to advanced meteor datasets, which contain a variety of directly observable features that may be used to indirectly constrain the physical properties of meteoroids. ML has successfully been applied in a number of recent meteor studies, including using ML to identify meteor showers \citep{Sugar2017, Pena-Asensio2025}, define meteor clusters \citep{Ashimbekova2025}, analyze images for meteor detections \citep{Gural2019, e-pena, Sicking2024}, and to match meteor spectra to their asteroidal counterparts \citep{DYAR2023115718}, among others. A few of those works - \cite{Sugar2017}, \cite{Ashimbekova2025}, and \cite{Pena-Asensio2025} - used similar techniques (i.e. dimensionality reduction and a clustering algorithm) to those employed here and thus will be summarized in the following paragraphs.

\cite{Sugar2017} was one of the first to use Density-Based Spatial Clustering of Applications with Noise \citep[DBSCAN;][]{ester_1996} to detect meteor showers. DBSCAN clusters samples based on a minimum number of points within a user-specified distance ($\epsilon$) and labels points falling outside of these density regions as noise. $\epsilon$ defines the maximum distance between points in a cluster. They used data from NASA's All-Sky Fireball Network and the Southern Ontario Meteor Network (SOMN).  After filtering out poor detections and merging both datasets to remove duplicates, they used as input 25,885 meteors in a six-dimensional parameter space. The first two dimensions were the cosine and sine of the meteor's solar longitude, describing its position in Earth's orbit at the time of observation. The next three dimensions pertained to the radiant's geocentric ecliptic latitude and Sun-centered ecliptic longitude. The final dimension was the meteor's geocentric velocity normalized to 72 $\mathrm{km\ s^{-1}}$. Each meteor was cloned 1000 times by assuming Gaussian-distributed measurement errors in ecliptic radiant and velocity. These 1000 clones were each processed with DBSCAN and master cluster lists were created. Clusters appearing in less than 100 iterations were discarded, while clusters sharing at least 50\% of their members were merged. With this method, 25 strong clusters were found and 6 weak clusters were identified, corresponding to the known meteor showers and shower complexes listed in the Meteor Data Center (MDC)\footnote[1]{\url{https://www.iaumeteordatacenter.org/}}. This study showed that unsupervised clustering algorithms can independently detect meteor showers from observational meteor data. 

\cite{Ashimbekova2025} applied DBSCAN to datasets from CAMS \citep{cams} and SonotaCo \citep{SonotaCo2009, SonotaCo2021} to identify meteor clusters. A meteor cluster is a set of meteors that share similar geocentric radiants and velocities and appear within a few seconds of each other \citep{Ashimbekova2025}. Meteor clusters form when a meteoroid fragments in space due to rotational pressure, collisions, or thermal stress \citep{Capek2022}, where the latter is considered the most likely cause. After fragmentation occurs, non-gravitational forces increase their separation, allowing the time since fragmentation to be derived from this distance. A single meteor camera network may detect clusters which fragment days before they enter Earth's atmosphere when fragment separation is still in the range of a few hundred $km$. However, to capture clusters with fragment separations in the thousands of $km$ range (equating to timescales of months since fragmentation), data from multiple geographically separated networks need to be used. This study combined CAMS data, taken in California, USA, with meteor detections from SonotaCo in Japan, spanning the time frame of 2007 January 1 to 2022 December 31. They used a four-dimensional vector consisting of time, geocentric velocity, the right ascension, and declination of the radiant. Each feature was normalized to the range of [-1, 1] except for time, which was scaled to [$-10^5, 10^5$], emphasizing its importance in finding meteor clusters. They found 85 initial clusters after applying DBSCAN, with 16 of those determined to be high-confidence clusters. They also found the $90^{th}$ percentile time difference between meteors in a cluster at the distances described was 8.0 seconds.

\cite{Pena-Asensio2025} analyzed how well the Hierarchical Density-Based Spatial Cluster of Applications with Noise \citep[HDBSCAN;][]{Campello2013} algorithm identified meteor showers using three different input vectors. HDBSCAN uses hierarchical clustering to group samples based on density, selects the most stable clusters, and classifies the remaining points as noise. Unlike DBSCAN, which requires both a minimum number of samples and an $\epsilon$ value as input, HDBSCAN only requires the minimum cluster size (\texttt{min\_cluster\_size}). The study analyzed 316,235 CAMS meteor detections, approximately 70\% of which were classified as sporadic meteors and the remaining were spread across 76 meteor showers. The three vectors tested were a geocentric parameter vector, a heliocentric orbit vector, and a vector from the CAMS lookup table. The CAMS lookup table vector, referred to in the paper as the LUTAB vector, consisted of the meteor's geocentric velocity, solar ecliptic longitude, and the radiant's Sun-centered ecliptic latitude and longitude. Clustering was evaluated using silhouette scores \citep{ROUSSEEUW198753}, Normalized Mutual Information \citep[NMI;][]{Vinh_2009}, and F1 scores. The F1 metric measures cluster quality relative to known classes by analyzing the precision and recall of cluster assignments. HDBSCAN identified 39 meteor showers when using a geocentric parameter vector and 30 when using a heliocentric orbit vector. Of these, 21 and 13, respectively, showed a strong match with CAMS classifications. The results were also dependent on the choice of minimum cluster size passed into HDBSCAN. This work showed that unsupervised clustering methods like HDBSCAN can perform statistically better than the CAMS look-up table method, although more work to physically validate this method remains to be done. 

Here, we apply an unsupervised clustering approach to classify meteors according to their inferred material strength based on directly observed characteristics. We use 13 features measured from low-light meteor cameras, such as velocity, magnitude, entry angle, heights, duration, energy received before ablation, atmospheric density, and trail length. These features are normalized and then processed with dimensionality reduction and clustering techniques to identify natural groupings in the data. We test numerous combinations of these techniques to determine the most effective algorithm. We use silhouette scores ($S$) to quantify cluster quality, while comparisons to existing models assess the physical realism of the clusters. Our goal is to develop a reliable and scalable method for classifying meteoroids that can handle the size and complexity of current and future datasets. Specifically, we aim to identify the combination of dimensionality reduction and clustering techniques that produces physically meaningful groupings. An additional requirement is that the resulting algorithm can be analytically implemented on future meteor datasets, or even single meteor events. We apply this methodology to nine well-studied meteor showers and examine clusters in orbital space to physically interpret the resulting classification scheme.

Before we begin, we acknowledge that multiple algorithm combinations could have been chosen as a result of our findings. This work demonstrates the approaches we tested, the algorithm we ultimately selected along with the rationale behind that choice, and our interpretation of the physical regimes revealed through the chosen method. We hope it will serve as a guide for future studies using clustering algorithms to uncover compositional patterns in meteor data. In Section~\ref{sec:methods}, we discuss our dataset along with the methodology behind the final normalization scheme, dimensionality reduction technique, and clustering algorithm. In Section~\ref{sec:rej_methods}, we summarize the combinations of algorithms we tested, the results of that testing, and the justification behind our choice of final model. Section~\ref{sec:results} outlines the results of our chosen approach. Finally, Section~\ref{sec:discussion} presents an application of this approach to well-studied meteor showers, patterns we found in orbital space, and the overall implications of our results. A \texttt{Python} implementation of the analytical model resulting from this work is made publicly available on GitHub (Section~\ref{code}). In the Appendix, the reader can find the mathematical formalism of the resulting model as well as all of the coefficients necessary to reproduce it.

\begin{table}[h]
\centering
\caption{Summary of meteor features.}
\label{features}
\begin{tabular}{ll}
\hline
\textbf{Symbol} & \textbf{Definition} \\
\hline
$E_\mathrm{beg}$ & Energy received before ablation (MJ/m\textsuperscript{2}) \\
$\rho_\mathrm{beg}$ & Atmospheric density at beginning height (kg/m$^3$) \\
$Ht_\mathrm{beg}$ & Beginning height of the meteor (km above WGS84 ellipsoid) \\
$Ht_\mathrm{end}$ & End height of the meteor (km above WGS84 ellipsoid) \\
$V_\mathrm{init}$ & Initial velocity (km/s) \\
$V_\mathrm{avg}$ & Average velocity (km/s) \\
$L_\mathrm{trail}$ & Trail length ($V_\mathrm{avg} \times t_\mathrm{dur}$, km) \\
$t_\mathrm{dur}$ & Duration of event (s) \\
$\theta_z$ & Zenith angle (90$^\circ$ -- Elevation angle, degrees) \\
$M_\mathrm{abs}$ & Peak magnitude normalized to 100 km \\
$F$ & F parameter: $(Ht_\mathrm{beg} - Ht_\mathrm{peak})/(Ht_\mathrm{beg} - Ht_\mathrm{end})$ \\
$m$ & Photometric mass (kg) \\
$a_\mathrm{decel}$ &
Deceleration $\left(V_\mathrm{init} - V_\mathrm{avg}\right)/t_\mathrm{dur}$ (km/s\textsuperscript{2}) \\
\hline
\end{tabular}

\vspace{2mm}
\footnotesize{Definitions are adapted from the GMN orbit data columns document.\footnote[2]{}}
\end{table}

\begin{figure}
    \centering
        \includegraphics[width=0.75\textwidth]{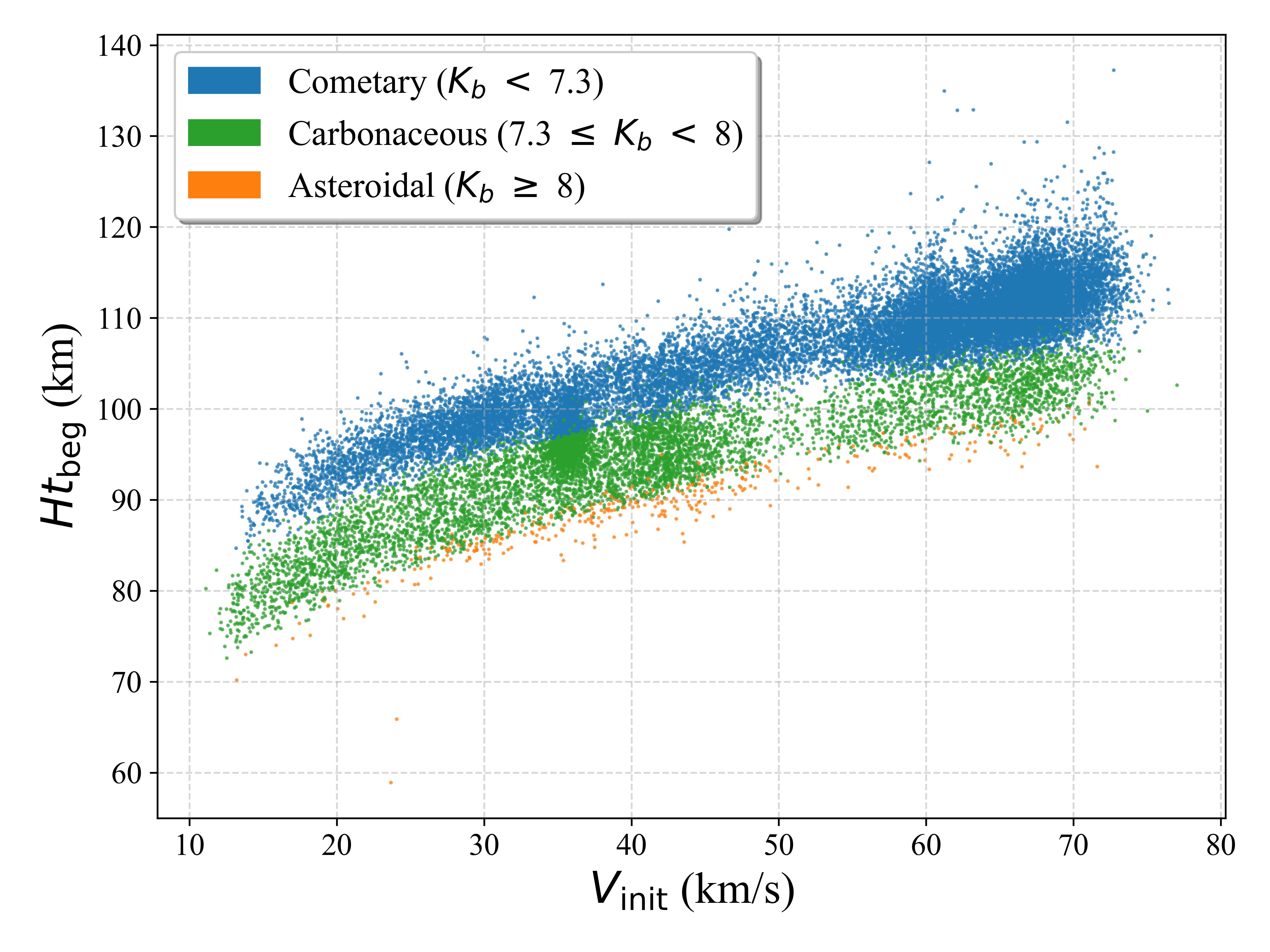}
    \caption{LO-CAMS meteor detections from 2023 with the best 50\% Median Fit Error (28,177 detections) color-coded based on their $K_{b}$ value calculated using Equation~\ref{k_b_87}. These groupings are used to to establish a reference for evaluating the performance of our clustering methods. Beginning height is on the y-axis and initial velocity is on the x-axis. Blue points are cometary meteors where $K_{b} < 7.3$, green points are carbonaceous meteors where $7.3 \leq K_{b} < 8.0$, and orange points are asteroidal meteors where $K_{b} \geq 8.0$. We implement a -0.10 correction to $K_{b}$ for GMN data as derived by \cite{Cordonnier_2024}. The silhouette score ($S=0.10$) was computed using the three features in Equation~\ref{k_b_87}.}
    \label{kb_ht_vel}
\end{figure}

\section{Methods} \label{sec:methods}

We used data collected in 2023 from the Lowell Observatory Cameras for All-sky Meteor Surveillance (LO-CAMS) to test and train algorithm combinations. LO-CAMS operates eight low-light meteor camera stations across the state of Arizona, each with six cameras providing all-sky coverage down to 30$^{\circ}$ elevation, and records high-resolution trajectory data for hundreds of meteors per night. Each LO-CAMS station utilizes Sony IMX291 CMOS IP cameras equipped with a 4mm \textit{f}/0.95 lens. Two stations, one in Holbrook, Arizona and one on the Lowell Observatory campus in Flagstaff, Arizona, route their six cameras to a single HP EliteDesk computer configured with Linux for multi-camera operation. The other six stations have each camera routed to a single Raspberry Pi3 board computer. The network is configured similarly to what is described in \cite{vida_2021}. LO-CAMS contributes observations to the CAMS\footnote[1]{\url{http://cams.seti.org/}} network along with the GMN. The GMN publishes its full dataset for public use\footnote[2]{\url{https://globalmeteornetwork.org/data}}. 

The 13 features we used are listed in Table~\ref{features}. $E_\mathrm{beg}$ is the energy the meteor received before ablating per unit of cross-sectional area in $MJ/m^2$ \citep{Borovicka_2007}. Beginning atmospheric density ($\rho_\mathrm{beg}$) is the density of the atmosphere where ablation began in $kg/m^3$. $E_\mathrm{beg}$ and $\rho_\mathrm{beg}$ were both computed for each meteor using the Western Meteor Python Library \citep[WMPL;][]{Vida_2019}. Beginning height ($Ht_{\mathrm{beg}}$) and ending height ($Ht_{\mathrm{end}}$) describe the heights in $km$ above the Earth's surface where ablation began and ended, respectively. Initial velocity ($V_{\mathrm{init}}$) is the meteor's apparent ground fixed velocity in $km/s$ upon entering the atmosphere. Average velocity ($V_{\mathrm{avg}}$) is the meteor's mean apparent ground fixed velocity in $km/s$ over the event's duration. Duration ($t_{\mathrm{dur}}$) is the time in seconds of the event's luminous period. Observed trail length ($L_{\mathrm{trail}}$) is the distance in $km$ the meteor traveled across the sky and is calculated as $V_{\mathrm{avg}}$ * $t_{\mathrm{dur}}$. Zenith angle ($\theta_{\mathrm{z}}$) is the angular distance of the meteor's apparent ground fixed radiant to the zenith. $\theta_{\mathrm{z}}$ is calculated from the dataset as 90° - elevation. $M_{\mathrm{abs}}$ describes the meteor's peak magnitude normalized to the range of 100 $km$. The lightcurve skewness parameter, $F$, describes the point along the meteor's trajectory where it experiences peak brightness. $F$ can be calculated as ($Ht_\mathrm{beg}$ – $Ht_\mathrm{peak}$)/($Ht_\mathrm{beg}$ - $Ht_\mathrm{end}$) and is dimensionless. $Ht_\mathrm{peak}$ is the meteor's height at peak brightness but is not a feature we used. The photometric mass ($m$) is the meteor's derived mass in $kg$. \cite{vida_2021} describes how this metric is calculated. Average deceleration ($a_{\mathrm{decel}}$) describes the rate in which the meteor's velocity changes in $km/s$\textsuperscript{2} throughout its luminous trajectory and can be calculated as $\left(V_\mathrm{init} - V_\mathrm{avg}\right)/t_\mathrm{dur}$. Prior to normalization, three features were kept in linear units ($t_{\mathrm{dur}}$, $M_{\mathrm{abs}}$, and $F$) and we computed the cosine of $\theta_{\mathrm{z}}$. The rest were scaled into logarithmic form. We note that several of these features are dependent upon or derived from other quantities in the feature set. The effect this has on our chosen algorithm is discussed in Section~\ref{subsubsec:dependencies}.

We employed a semi-qualitative approach to finding clusters based on canonical separations in meteor data. Previous studies have shown that softer, cometary meteors begin ablating higher in the atmosphere than harder, asteroidal meteors \citep{Ceplecha_and_McCrosky_1976, Koten_2004}. This separation is shown in our data in Figure~\ref{kb_ht_vel}, where low density cometary meteors have higher beginning heights than denser, carbonaceous and asteroidal meteors. We used these separations to assess whether the clusters produced by different models were physically realistic. As such, the same $Ht_{\mathrm{beg}}$ vs. $V_{\mathrm{init}}$ visualization was employed to qualitatively analyze the groupings produced by various algorithms. We refer to clusters resembling those seen in Figure~\ref{kb_ht_vel} as realistic results and those which are visually unlike those groupings as unrealistic results. Silhouette scores were used as a quantitative way to evaluate cluster quality. See Section~\ref{subsec:sil_score} for a description of the silhouette score metric.

Several combinations of normalization, dimensionality reduction, and clustering techniques were tested, but many produced results that were either not physically meaningful or not generalizable for future datasets. Our general workflow consisted of selecting and normalizing our dataset, conducting dimensionality reduction, and then applying a clustering algorithm. We used functions in the Python-based \texttt{scikit-learn} \citep{scikit-learn} toolkit. Here, we discuss our methods in terms of the final algorithm we selected: Factor Analysis (FA) for dimensionality reduction and Gaussian Mixture Model (GMM) for clustering.    

\begin{figure}
    \centering
        \includegraphics[width=\textwidth]{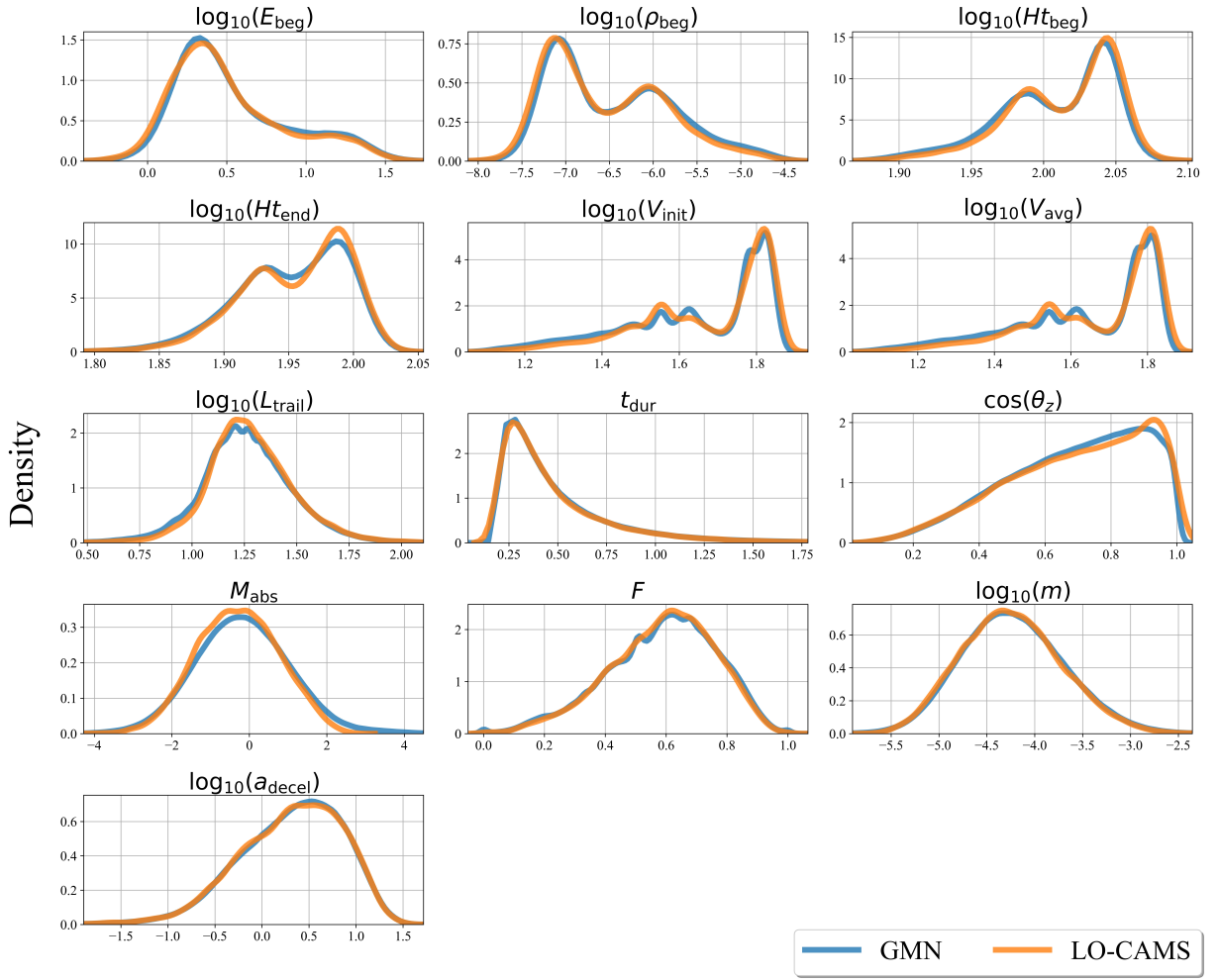}
    \caption{Histograms for each feature showing that the distributions in LO-CAMS 2023 data matches that of the entire GMN dataset. GMN data is in blue and LO-CAMS in orange. The GMN dataset was subject to the same median fit error cutoffs applied to the LO-CAMS dataset (Section~\ref{subsec:data}) and contains 825,864 data points. The Y-axes show the normalized probability density for each feature, rather than raw counts. As a result, heights can exceed unity in the case of small bin widths. The X-axis shows scaled values; see Section~\ref{sec:methods} for details on how each feature was scaled.}
    \label{feature_hist_overlay}
\end{figure}

\subsection{Data Collection and Feature Construction} \label{subsec:data}

The LO-CAMS 2023 dataset consisted of 56,407 valid meteor events. A valid meteor was an event that did not have any $NaN$ values and had values that made physical sense (i.e. some entries had negative mass values). We then used the event's median fit error to determine precisely measured events. The median fit error is a calculation of the median angular trajectory fit errors in arc seconds. We used the LO-CAMS events with the lowest 50\% median fit errors, giving us 28,177 meteor events for the basis of deriving clusters. The 50\% threshold for median fit errors for this data set was $59.46^{\prime\prime}$. We found that factor structure and cluster quality were reasonably consistent between 10\% and 90\%. Thus, we chose 50\% as a trade-off between data quality and quantity. Although the GMN includes measurement uncertainties for all reported quantities, we used the nominal measured values and did not incorporate the measurement errors into the features. 

Clustering algorithms are sensitive to the distributions of the features used. In practice, whether a feature is approximately normal, skewed, or contains structure such as peaks and valleys can influence how the algorithm identifies groups. Therefore, an important step was to ensure that the training set we used was representative of the larger population. Figure~\ref{feature_hist_overlay} shows histograms of each feature for the entire GMN dataset observed between 2018 December 10 through 2025 May 16 (in blue) and for the LO-CAMS data (in orange) demonstrating that their distributions are highly similar. Although this is the case, the LO-CAMS data may be subject to latitude-dependent selection effects since it was collected around a limited region on Earth. The GMN dataset represented nearly all of the observations conducted by the network worldwide since its inception. It consisted of 1,652,276 valid meteor events, 825,864 of which remained after applying the median fit error cut off used for the LO-CAMS data. We attempted to use a subset of the entire GMN data to train our ML algorithm, on the order of 200,000 to 300,000 meteor events, but found that too much data created unrealistic models. We used a set of well-studied showers from this dataset to physically interpret the clusters our method identified. The details of that analysis are discussed in Section~\ref{subsec:showers}.

\subsection{Data Normalization} \label{subsec:preprocess}
Before applying ML methods, we normalized the dataset using the \texttt{MinMaxScaler} function to ensure equal weighting across all features. All values were scaled within the range [-1, 1] instead of the default range [0, 1] because we found the [-1, 1] range gave us more pronounced factor weightings when we applied Factor Analysis. The normalization step is necessary to ensure each feature has equal relevance, as features on larger scales can disproportionately influence the results. 

We tested several normalization schemes from the \texttt{scikit-learn} package, including \texttt{StandardScaler}, \texttt{RobustScaler} and \texttt{normalize}. \texttt{StandardScaler} and  \texttt{RobustScaler} both produced clusters with  $S$ comparable to those obtained with \texttt{MinMaxScaler}. However, the dimensionality reduction step after using those normalization schemes produced results for each new axis that contained a large number of strongly contributing features, making interpretation more complex. \texttt{MinMaxScaler} produced factors or axes with fewer strongly contributing features, simplifying interpretation. The \texttt{normalize} function produced unrealistic clusters on the $Ht_{\mathrm{beg}}$ vs. $V_{\mathrm{init}}$ plot and was discarded.

\subsection{Factor Analysis} \label{subsec:fa}
To reduce dimensionality and highlight key patterns in the data, we applied Factor Analysis \citep{Spearman1904, Thurstone1935} to the normalized feature set. FA models the data as a linear combination of latent variables (the factors) and Gaussian noise. It focuses on identifying the unobserved factors governing the causal structure of the data. FA is widely used in the behavioral and social sciences to infer the latent drivers of covariance among related variables \citep{miller2007handbook} and can be applied to any domain involving high-dimensional, correlated data. The difference between principal component analysis \citep[PCA;][]{Jolliffe_2002} and FA is that PCA attempts to maximize variance within the data while FA models the covariance among features to identify underlying latent dimensions. In terms of mathematical formalism, following \cite{barberBRML2012}, the $i^{th}$ event of normalized dataset $X'$ with $n$ observed features can be defined as:

    \begin{equation}
        \label{vector}
        x'_i = \{x'_{i,1}, x'_{i,2}, \ldots, x'_{i,n}\}
    \end{equation}

Under FA, each observation is modeled as a function of $j$ unobserved factors:

    \begin{equation}
        \label{factorVector}
        x'_i = (w_1 f_{i,1}) + (w_2 f_{i,2}) + \ldots + (w_j f_{i,j}) + \mu + \epsilon_{i} 
    \end{equation}

Where $w_{n}$ are the factor loadings (weights) for each feature, $f_{i,n}$ are the latent factors, $\mu$ is the mean vector of the normalized features (referred to as a bias term), and $\epsilon_i$ is a noise term. $\epsilon$ describes variation due to measurement noise and unmeasured variables not taken into account by the common factors. $\epsilon$ follows a multivariate normal distribution with zero mean and covariance $\Psi$:

    \begin{equation}
        \label{epsilon}
        \epsilon_i \sim \mathcal{N}(0, \Psi)
    \end{equation}

The definition of $\Psi$ dictates the difference between PCA and FA. If $\Psi = \sigma^2I$, where I is the identity matrix, PCA is recovered. In standard FA, $\Psi = diag(\psi_1, \psi_2, ..., \psi_n)$, allowing each feature to have its own unique variance. In matrix notation, Equation~\ref{factorVector} takes the form:

    \begin{equation}
        \label{Matrix}
        X' = FW^{\top}+ M + E 
    \end{equation}  

Where $X'$ is the normalized ($m \times n$) dataset containing $m$ events, $F$ is the ($m \times j$) matrix of factor scores, $W$ is the ($n \times j$) matrix of factor loadings, $M$ is the matrix of bias terms, and $E$ is the matrix of noise terms. Solving for the factor score matrix, $F$, yields a dimensionally reduced representation of the data in $j$-dimensional factor space. This solution is obtained by computing a linear estimator that projects the normalized observations onto the factor loadings while accounting for feature-specific noise. The resulting factor scores $F$ are subsequently used as input to the GMM. The mathematical formalism implemented in our analytical model to compute $F$ is provided in Appendix~\ref{subsec:fa_model}.

\subsection{Gaussian Mixture Model} \label{subsec:gmm}
After normalizing the data and reducing its dimensionality with FA, we applied the \texttt{scikit-learn} \\ \texttt{GaussianMixture} implementation of a GMM to identify potential groupings. The GMM models the observed data as a mixture of $k$ multivariate Gaussian components, each representing a cluster of meteors defined by a mixture weight ($\pi_{k}$), mean vector ($\mu_{k}$), and covariance matrix ($\Sigma_{k}$). The number of clusters ($k$) is passed as input by the user. For any given event's factor vector, $f_{m}$, the probability density function (PDF) for cluster $k$ under a $j$ factor model is defined as:

\begin{equation}
    \mathcal{N}(f_{m}|\mu_{k}, \Sigma_{k}) = \frac{1}{\left(2\pi\right)^{\frac{j}{2}}|\Sigma_{k}|^{\frac{1}{2}}}exp^{\left[-\frac{1}{2}\left(f_{m}-\mu_{k}\right)^{\top}\Sigma_{k}^-1\left(f_{m}-\mu_{k}\right)\right]}
    \label{equ:methods_gmm1}
\end{equation}

The likelihood of observing $f_{m}$ under the full GMM is specified by: 

\begin{equation}
    p\left(f_{m}|\Theta\right) = \Sigma_{k=1}^{K}\pi_{k}\mathcal{N}(f_{m}|\mu_{k}, \Sigma_{k})
    \label{equ:methods_gmm2}
\end{equation}

Where $\Theta = \{\pi_{k}, \mu_{k}, \Sigma_{k}\}$ are the model parameters. The \texttt{GaussianMixture} implementation initializes these values by default using a K-Means style initialization. Model parameters are then optimized using the Expectation–Maximization (EM) algorithm \citep{Dempster_1977}, which iteratively increases the incomplete-data log-likelihood of the dataset:

\begin{equation}
    \Sigma_{m=1}^{N}\log p\left(f_{m}|\Theta\right) 
\label{equ:methods_gmm3}
\end{equation}

Log-likelihood is the log of the probability of the observed data under the model, given its fitted parameters. A higher score indicates a better fit model. Because cluster membership is previously unknown, EM alternates between estimating posterior cluster probabilities ($\gamma_{mk}$) and updating the model parameters. In practice, this corresponds to updating parameters as if cluster assignments were known, but with each event weighted by its posterior probability of belonging to each cluster. This process is iterated until successive increases in the incomplete-data log-likelihood fall below a specified tolerance.

In the Expectation step, the posterior probability that event $m$ belongs to cluster $k$ is computed as:

\begin{equation}
    \gamma_{mk} = p(z_m = k \mid f_m, \Theta) = \frac{\pi_k \mathcal{N}(f_m \mid\mu_k, \Sigma_k)}{\sum_{l=1}^{K}\pi_l \mathcal{N}(f_m \mid \mu_l, \Sigma_l)}
\label{equ:methods_gmm4}
\end{equation}

In the Maximization step, model parameters are updated by incorporating the posterior cluster probabilities obtained in the Expectation step (Equation~\ref{equ:methods_gmm4}). Each event contributes to the parameter estimates of all clusters in proportion to its probability of belonging to each cluster. Parameter updates are obtained by maximizing the expected complete-data log-likelihood, which is accomplished by replacing the unknown cluster assignments with their posterior probabilities when computing the mixture weights, cluster means, and covariance matrices. This corresponds to computing probability-weighted averages of the factor vectors for each cluster. Following each update, the incomplete-data log-likelihood of the observed factor vectors is re-evaluated, and the Expectation and Maximization steps are iterated until this log-likelihood converges. Once completed, this procedure returns hard cluster assignments along with the posterior probability of membership in each cluster. The ideal number of clusters for the GMM was determined by examining how silhouette scores $S$ and Bayesian Information Criterion \citep[BIC;][]{Cherkassky_2003, Zou_2007} changed as the number of clusters was increased. BIC is commonly used with GMMs and is discussed in Section~\ref{subsec:rej_clustering}.

\subsection{Silhouette Score} \label{subsec:sil_score}
The silhouette score, introduced in \cite{ROUSSEEUW198753}, serves as a metric for validating cluster quality. It operates by calculating a score for each individual data point $i$ based on two values: its cohesion and its separation. Cohesion, $a(i)$, is the mean distance between point $i$ and all other points in its own cluster. Separation, $b(i)$, is the mean distance from point $i$ to all points in the single nearest neighboring cluster.

These two values are used to compute the silhouette score for that point, $S(i)$, as defined in Equation~\ref{silhouette}:

\begin{equation}
\label{silhouette}
S(i) = \frac{b(i) - a(i)}{\max\{a(i), b(i)\}}
\end{equation}

The overall silhouette score for the entire dataset is the mean of the $S(i)$ values for all points. The score ranges from $-1$ to $1$. A score closer to $1$ indicates that points are well-matched to their assigned clusters and distinct from neighboring clusters (i.e. $a(i)$ is much smaller than $b(i)$). A score closer to $-1$ suggests points are likely misclassified, as they are, on average, closer to a neighboring cluster than their own (i.e. $b(i)$ is smaller than $a(i)$). We used the \texttt{scikit-learn's silhouette\_score} function for this computation.

\section{Tested and Rejected Methods} \label{sec:rej_methods} 
 
We conducted a series of trials using three dimensionality reduction techniques and six clustering algorithms to determine the optimal workflow. Initial trials informed the number of components to use in the dimensionality reduction step. For clustering, we examined how $S$ evolved as we increased the number of clusters while comparing the groupings to Figure~\ref{kb_ht_vel}. This testing ultimately informed our selection of dimensionality reduction and clustering techniques, as well as the number of clusters we chose to analyze.   

\subsection{Dimensionality Reduction Analysis} \label{subsec:rej_dimReduc}
PCA is relatively straightforward because the method will return the amount of variance explained by each component. A common guideline is to retain enough components to capture 95\% of the total variance. Applying PCA to the LO-CAMS data revealed that 5 components captured 95\% of the data's variance, guiding the number of PCA components used for our clustering experiments. 

FA, on the other hand, is not quite as clear-cut. One method of determining an appropriate number of factors is to analyze how the model's log-likelihood changes as the number of factors increases. We found that the log-likelihood stopped increasing at 6 factors, indicating a maximum number of factors to use in our model. Although a larger number of factors would provide a better model, a smaller number of factors makes the final model less complex and easier to interpret. Motivated by the need to find a trade off between the smallest number of factors and the best fit model, we assessed clustering performance using between 2 and 6 factors.

We used a GMM for this initial assessment because some features exhibited approximately normal distributions while others contained structure (Figure~\ref{feature_hist_overlay}). The probabilistic nature of GMM makes it sensitive to structure in the data and allowed us to evaluate factor selection even when some features were not normally distributed. In the next section, we expand beyond GMM and evaluate how several clustering algorithms performed using these factor and component sets. Using a 3 cluster GMM, we found that $S$ decreased as we increased the number of factors. 2 factors resulted in the highest $S=0.41$, but clusters were blended and not physically realistic in $Ht_{\mathrm{beg}}$ vs. $V_{\mathrm{init}}$ space. 3 factors produced the next highest $S=0.29$ and produced clusters in line with Figure~\ref{kb_ht_vel}. The highest $S$ for the remaining tests (4, 5, and 6 factors) was $S=0.17$. We chose to move forward using 3 factors because it struck a balance between a small number of factors, the highest $S$, and realistic results.  

A requirement of any resulting algorithm was that the model can be analytically applied to future datasets and individual events. To that end, we avoided dimensionality reduction techniques that would be difficult to implement on new data. At the same time, we wanted to explore the ability of other methods to find patterns in meteor data. One such approach was Uniform Manifold Approximation and Projection (UMAP) \citep{mcinnes2020}, which is a manifold learning technique that reduces dimensionality while retaining the non-linear structure of the data. Although UMAP can be effective at uncovering complex patterns, the resulting embeddings are often data-dependent, making it difficult to apply the transformations to future data. 

\begin{figure}
    \centering
        \includegraphics[width=\linewidth]{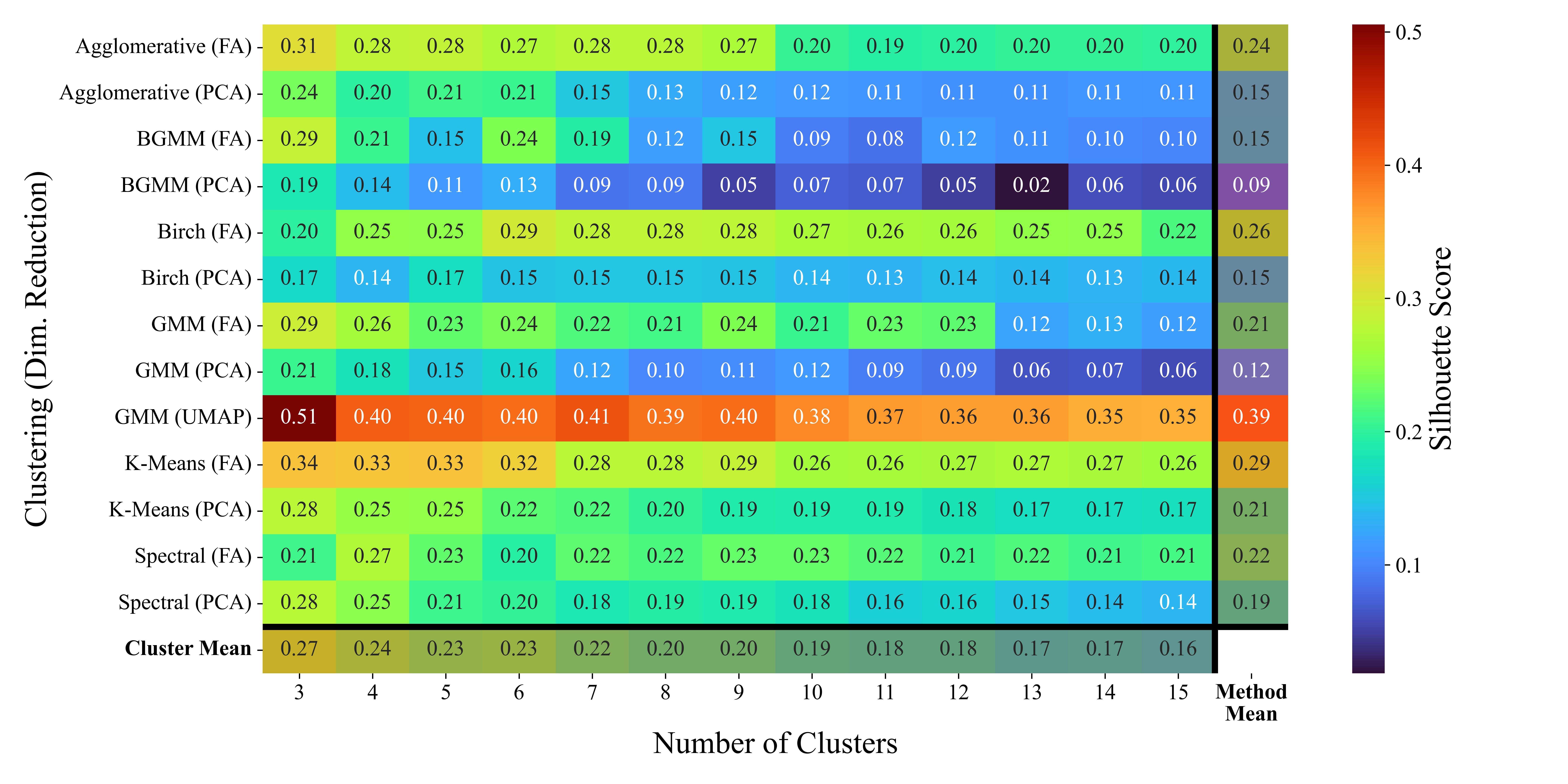}
    \caption{Heatmap of how $S$ changes as we increase the number of clusters for workflows using different dimensionality reduction techniques (FA, PCA, UMAP) and clustering methods (Agglomerative, BGMM, Birch, GMM, K-Means, and Spectral). The last row displays the mean silhouette score for each number of clusters and the last column shows the mean score for each clustering method.}
    \label{fig:silhouette_scores}
\end{figure}

\subsection{Clustering Method Analysis} \label{subsec:rej_clustering}

We focused on clustering algorithms which required minimal tuning of hyper-parameters. The six clustering algorithms we tested were: \texttt{AgglomerativeClustering}, \texttt{BayesianGaussianMixture} (BGMM), \texttt{Birch}, \texttt{GaussianMixture}, \texttt{KMeans}, and \texttt{SpectralClustering}. They are all implemented in the \texttt{sklearn.cluster} and \texttt{sklearn.mixture} packages. We observed a grand mean $S=0.20$ among all methods tested. Comparing this with $S=0.10$ from $K_{b}$ indicates that an expanded feature set analyzed with machine learning methods can enhance the amount of information we are able to extract from meteor data. Figure~\ref{fig:silhouette_scores} shows a heatmap of $S$ for each combination of dimensionality reduction techniques and clustering methods as a function of the number of clusters, revealing an overall decrease in $S$ as the number of clusters increases. FA generally produced higher $S$ than PCA. 

\begin{figure}
    \centering
    % ------------------- Top Row (2 images) -------------------
    \begin{subfigure}[t]{0.30\textwidth}
        \centering
            \includegraphics[width=\linewidth]{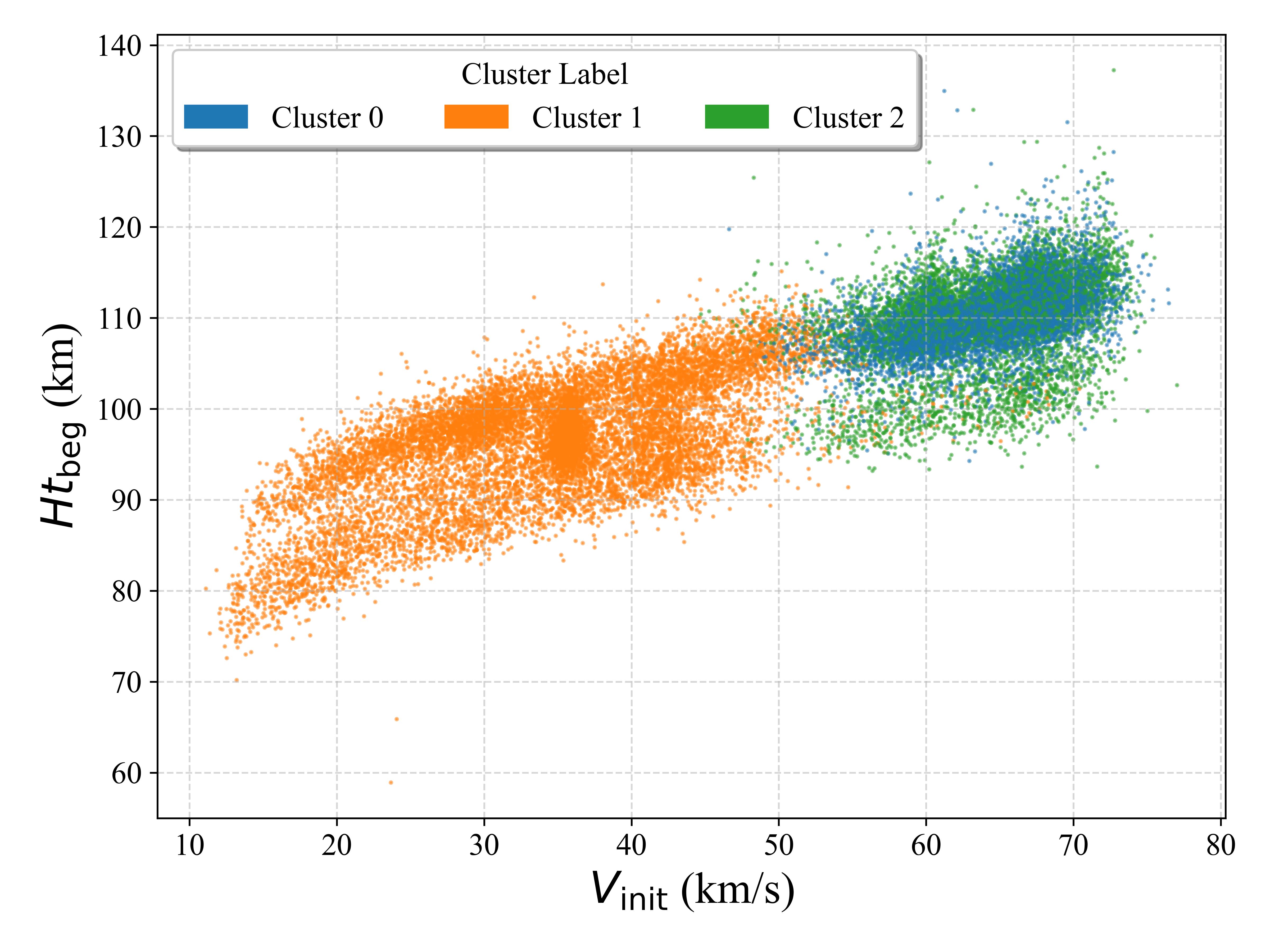}
        \caption{UMAP + GMM\\$S=0.51$}
        \label{rej_umapGmm}
    \end{subfigure}
    \hfill
    \begin{subfigure}[t]{0.30\textwidth}
        \centering
            \includegraphics[width=\linewidth]{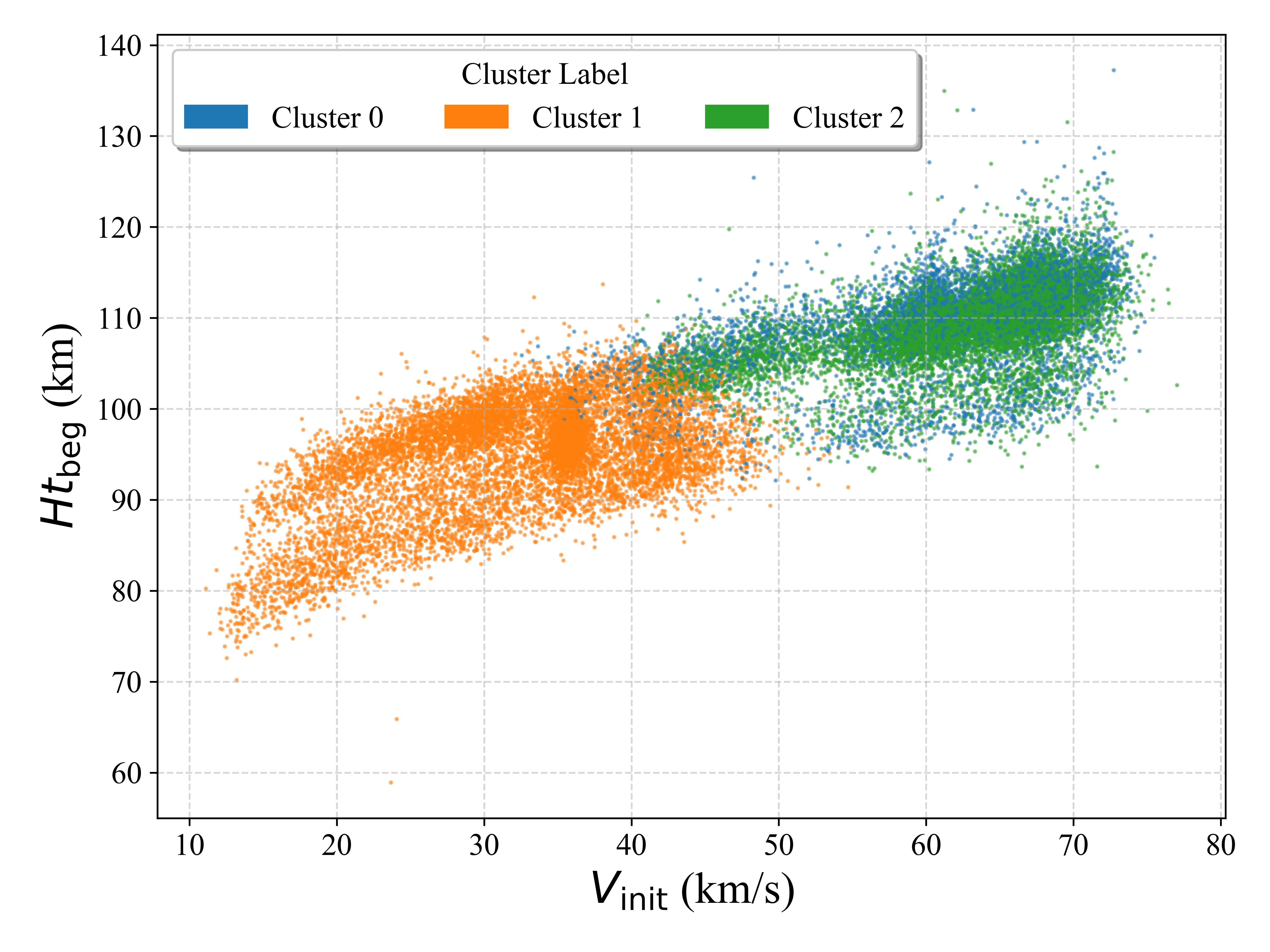}
        \caption{PCA + Spectral\\$S=0.28$}
        \label{rej_pcaSpec}
    \end{subfigure}
    \hfill
    \begin{subfigure}[t]{0.30\textwidth}
        \centering
            \includegraphics[width=\linewidth]{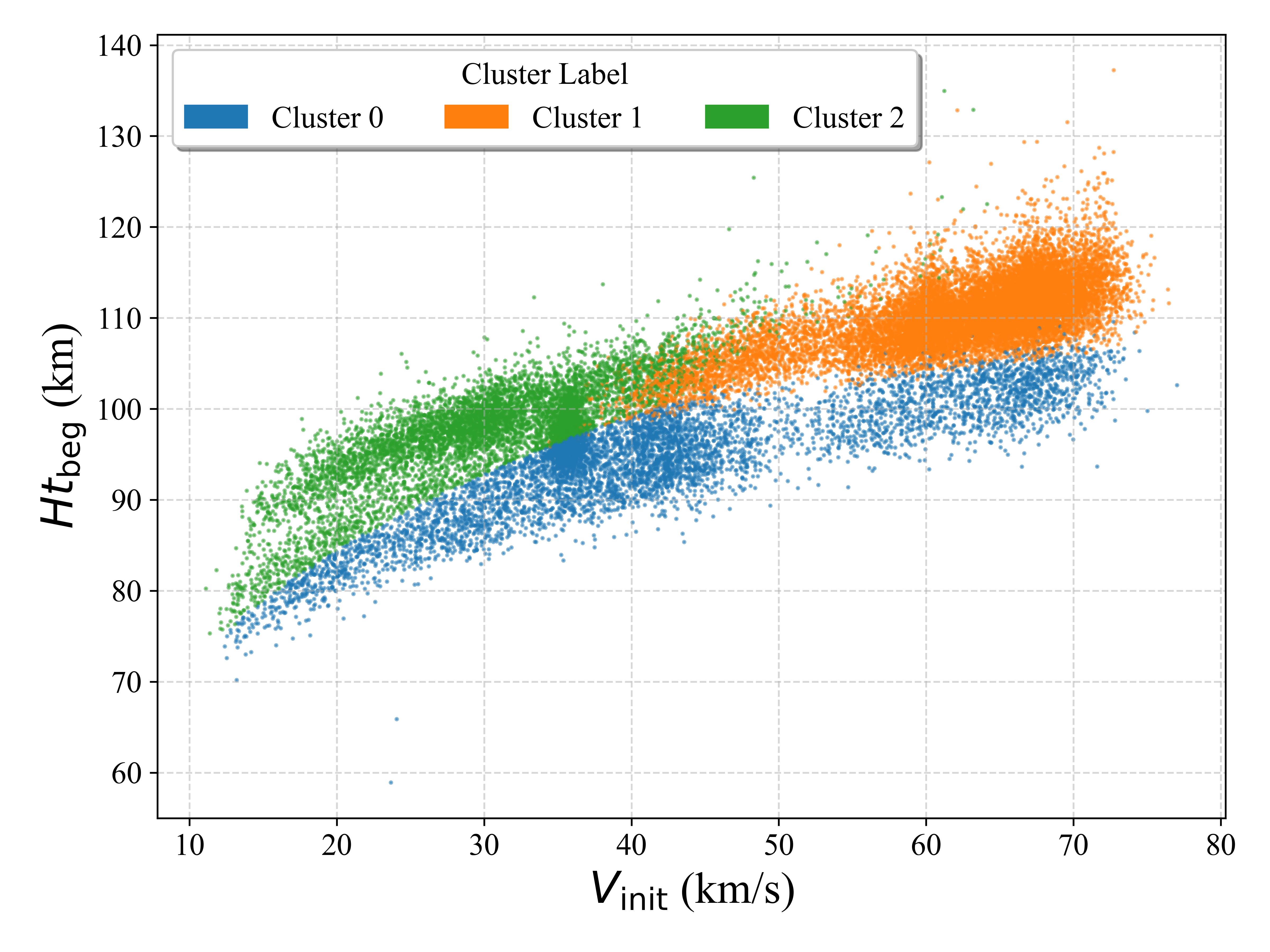}
        \caption{FA + KMEANS\\$S=0.34$}
        \label{rej_faKmeans}
    \end{subfigure}

    \vspace{1em} % space between rows

    % ------------------- Bottom Row (3 images) -------------------
    \begin{subfigure}[t]{0.30\textwidth}
        \centering
            \includegraphics[width=\linewidth]{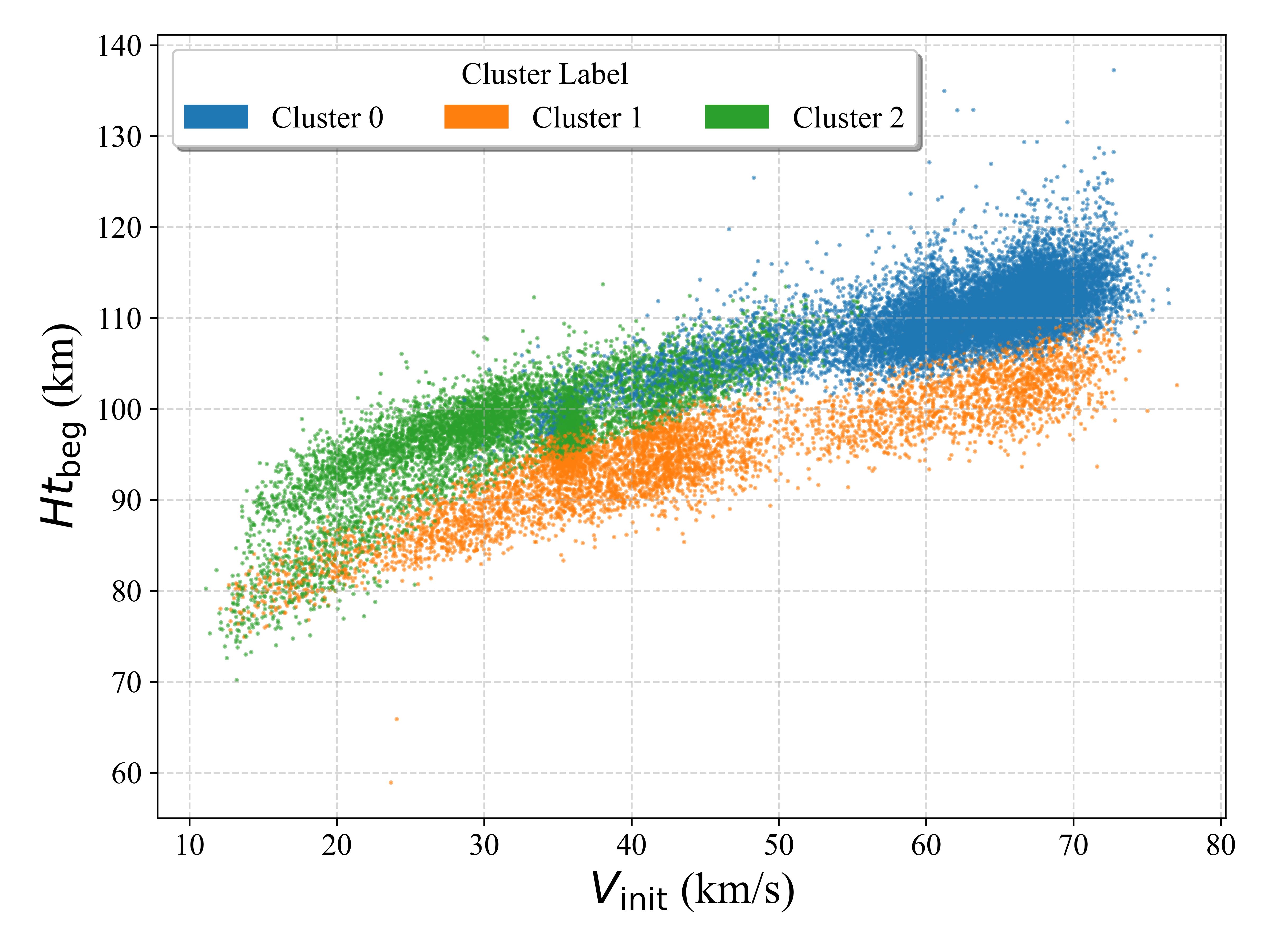}
        \caption{FA + Agglomerative\\$S=0.31$}
        \label{rej_faAgglom}
    \end{subfigure}
    \hfill
    \begin{subfigure}[t]{0.30\textwidth}
        \centering

                \includegraphics[width=\linewidth]{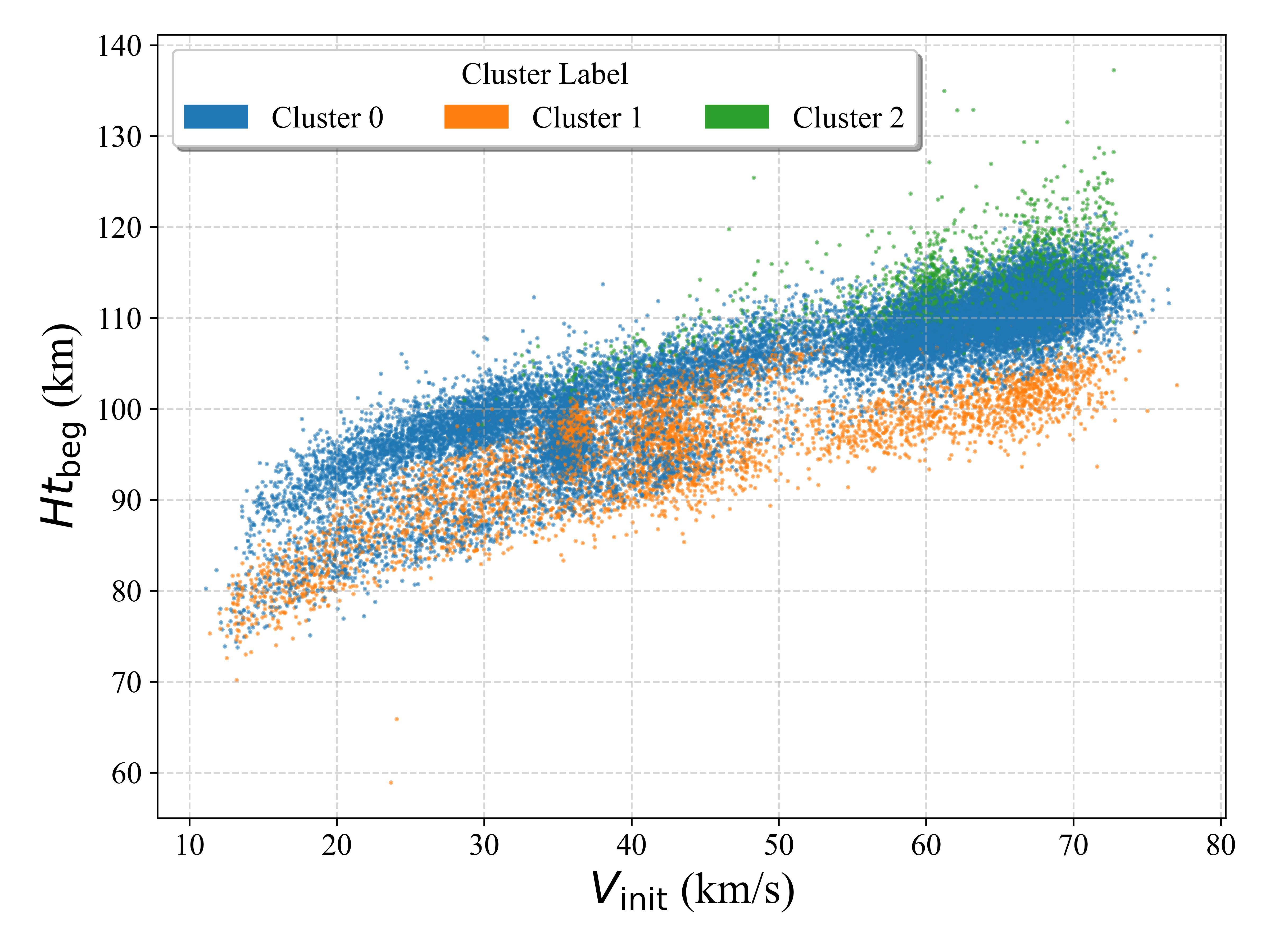}
        \caption{FA + Birch\\$S=0.20$}
        \label{rej_faBirch}
    \end{subfigure}
    \hfill
    \begin{subfigure}[t]{0.30\textwidth}
        \centering
            \includegraphics[width=\linewidth]{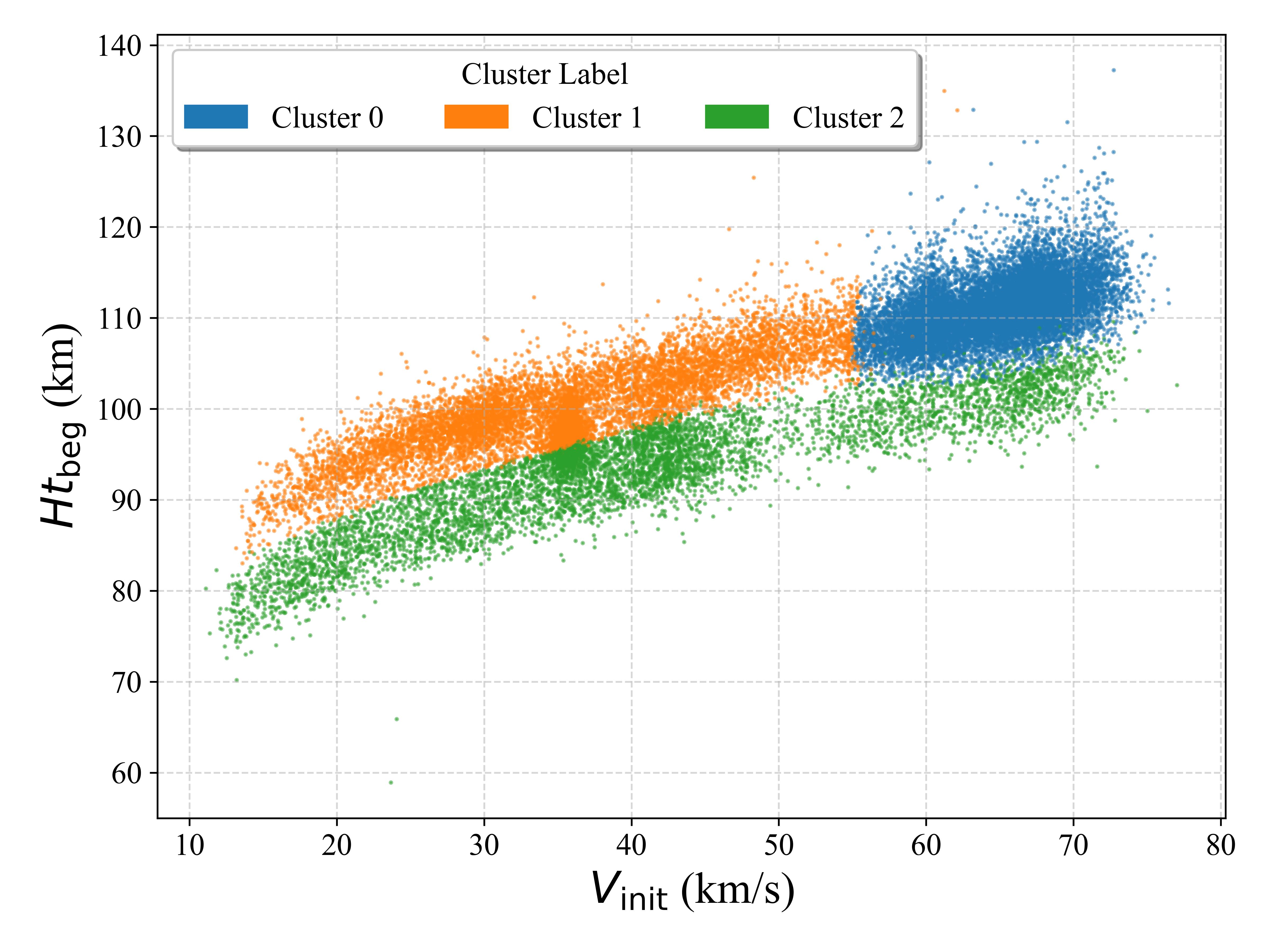}
        \caption{FA + BGMM\\$S=0.29$}
        \label{rej_faBgmm}
    \end{subfigure}
    \caption{$Ht_\mathrm{beg}$ vs. $V_\mathrm{init}$ plots of rejected methods. These methods exhibited unrealistic behavior, such as diagonal or vertical separations and cluster overlap. We compared the groupings found by different combinations of dimensionality reduction techniques and clustering algorithms to the groupings revealed in this space by $K_{b}$ (Figure~\ref{kb_ht_vel}). The FA + BGMM (\ref{rej_faBgmm}) 3 cluster results are very similar to what was produced by our final model (FA + GMM, Figure~\ref{3_cluster_result}), however this combination produced weaker results as we increased the number of clusters in the model.}
    \label{fig:rej_methods}
\end{figure}

The UMAP-GMM combination produced the highest mean silhouette score ($S=0.39$) across all cluster values. However, this combination was ultimately rejected for two reasons. First, inspection of its $Ht_\mathrm{beg}$ vs. $V_\mathrm{init}$ plot revealed unrealistic results (Figure~\ref{rej_umapGmm}). Many of the methods we tested showed similar issues in this space, including most data points being assigned to a single cluster (see Section~\ref{subsec:HDBSCAN} on HDBSCAN), diagonal or vertical separations that didn't make physical sense (Figure~\ref{rej_umapGmm} and \ref{rej_pcaSpec}), and/or cluster blending (Figure~\ref{rej_pcaSpec}-\ref{rej_faBirch}). We will discuss the inherent bias presented by using semi-qualitative approach in Section~\ref{subsec:bias}. Second, UMAP is a manifold learning technique and therefore cannot be easily implemented in future applications without rerunning the model on a large data set. 

PCA-Spectral (Figure~\ref{rej_pcaSpec}) is provided as another example of diagonal/vertical separations we deemed physically unrealistic. FA-KMeans produced the second highest mean silhouette score ($S=0.29$) and generally realistic clusters, although we observed poor segregation of cometary and carbonaceous groups at low velocities (Figure~\ref{rej_faKmeans}). The FA-Agglomerative combination yielded a higher mean silhouette score ($S=0.24$) than FA-GMM ($S=0.21$). This combination was rejected because the high- and low-velocity cometary clusters were highly blended, and also due to poor segregation of cometary and carbonaceous groups at low velocities (lower left points in Figure~\ref{rej_faAgglom}). FA-Birch (Figure~\ref{rej_faBirch}) was another combination that exhibited a higher mean silhouette score than FA-GMM ($S=0.26$). While its 3 cluster $S=0.20$ is the lowest of the trials presented in Figure~\ref{fig:rej_methods}, FA-Birch generally exhibited higher $S$ than FA-GMM in trials containing $\geq 4$ clusters. FA-Birch was rejected due to a high degree of cluster blending. Similar cluster blending was observed throughout all trials of this method. The FA-BGMM combination produced clustering results comparable to FA-GMM (Figure~\ref{rej_faBgmm}), but the method's mean $S=0.15$ was less than the mean $S=0.21$ for FA-GMM.

We selected the FA-GMM algorithm as our final model primarily because it offered a balance between silhouette score and physically realistic clusters, as evident from the $Ht_\mathrm{beg}$ vs. $V_\mathrm{init}$ plots (See Section~\ref{sec:results}). A complementary reason for selecting the GMM is that it is more analytically tractable, once the model has been trained, than the other clustering methods we explored. It had the third highest mean $S$ of all methods tested. The number of clusters used to guide subsequent analyses with GMM was chosen based on the $S$ trend together with BIC. BIC evaluates model quality by balancing goodness of fit with model complexity, penalizing models with more parameters to discourage overfitting, where lower BIC scores indicate a better model. BIC is automatically returned by the \texttt{GaussianMixture} function. 

\begin{figure}
    \centering
        \includegraphics[width=0.7\linewidth]{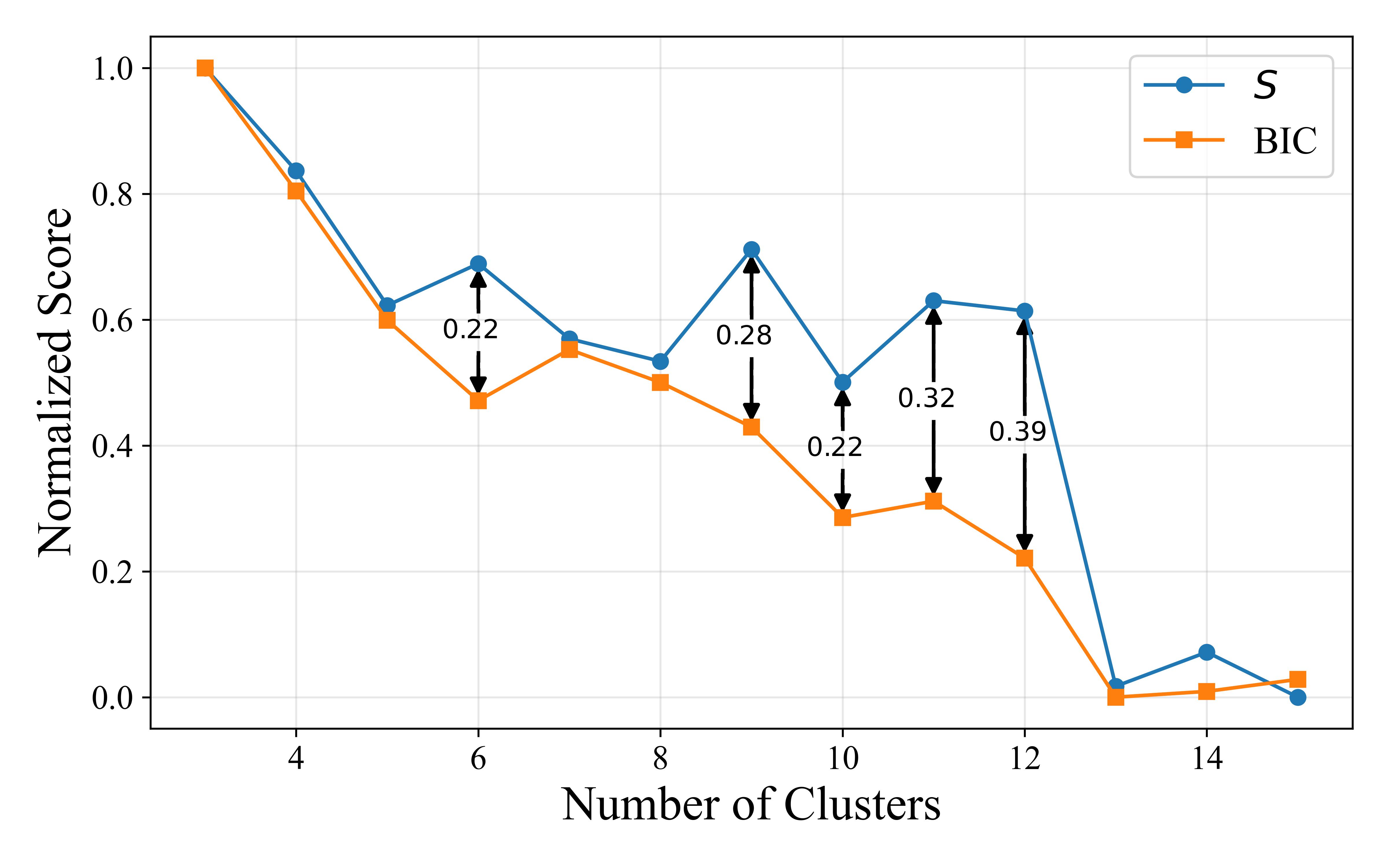}
    \caption{The separation between normalized $S$ and BIC ($\Delta$) varies with cluster count, where noticeable $\Delta$s are found at 6 and 9-12 clusters. The combined behavior of these scores guided the selection of the number of clusters used in the GMM. Scores are normalized between the range $[0,1]$ to enable direct comparison since both metrics are defined on different scales. $S$ is shown in blue and BIC in orange. Annotations indicate the $\Delta$ between $S$ and BIC for the given cluster values.}
    \label{fig:bic_s}
\end{figure}

Figure~\ref{fig:bic_s} illustrates how $S$ and BIC vary with the number of clusters. The separation between these two scores ($\Delta$) highlights cluster values where cluster resolution ($S$) and model complexity (BIC) may be more favorable and informed the number of clusters to use. The 3 cluster GMM exhibited the highest $S$ and BIC, however it represented broad physical groupings (e.g. cometary, carbonaceous, asteroidal) and provided realistic results when compared to Figure~\ref{kb_ht_vel}. The two metrics trended together until 6 clusters, which showed the first divergence between $S$ and BIC ($\Delta=0.22$). Both metrics again trended together until 9 clusters was reached. The largest $\Delta$ between 7 and 15 clusters was at 12 ($\Delta=0.39$), 11 ($\Delta=0.32$), and 9 ($\Delta=0.28$) clusters. We selected 11 clusters over 12 because the 12 cluster solution exhibited more blending.

\subsection{HDBSCAN} \label{subsec:HDBSCAN}

Based on the success of DBSCAN and HDBSCAN in previous meteor studies \citep{Sugar2017, Ashimbekova2025, Pena-Asensio2025}, we evaluated how HDBSCAN performed on our data. We selected HDBSCAN because it offers similar clustering capabilities to DBSCAN but with fewer required hyper-parameters and is better suited to noisy data. Two dimensionality reduction techniques (FA and PCA) were used while UMAP was excluded since it does not meet our requirement of being generalizable to future data. 

\begin{figure}
    \centering
    \begin{subfigure}[t]{0.48\textwidth}
        \centering
            \includegraphics[width=\linewidth]{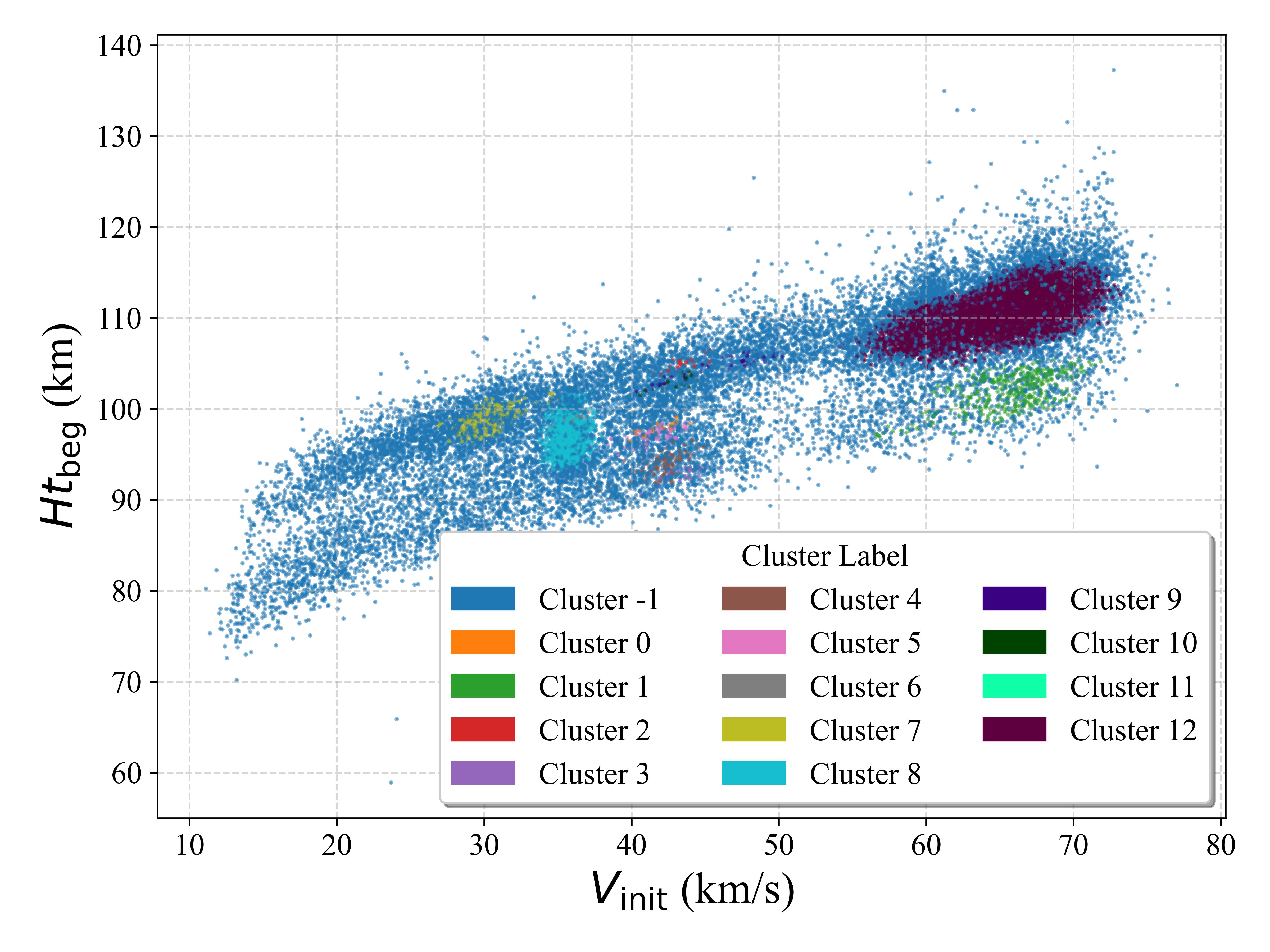}
        \caption{\centering FA + \texttt{min\_cluster\_size} = 20\\$S=-0.32$}
        \label{hdbscan_a}
    \end{subfigure}
    \hfill
    \begin{subfigure}[t]{0.48\textwidth}
        \centering
            \includegraphics[width=\linewidth]{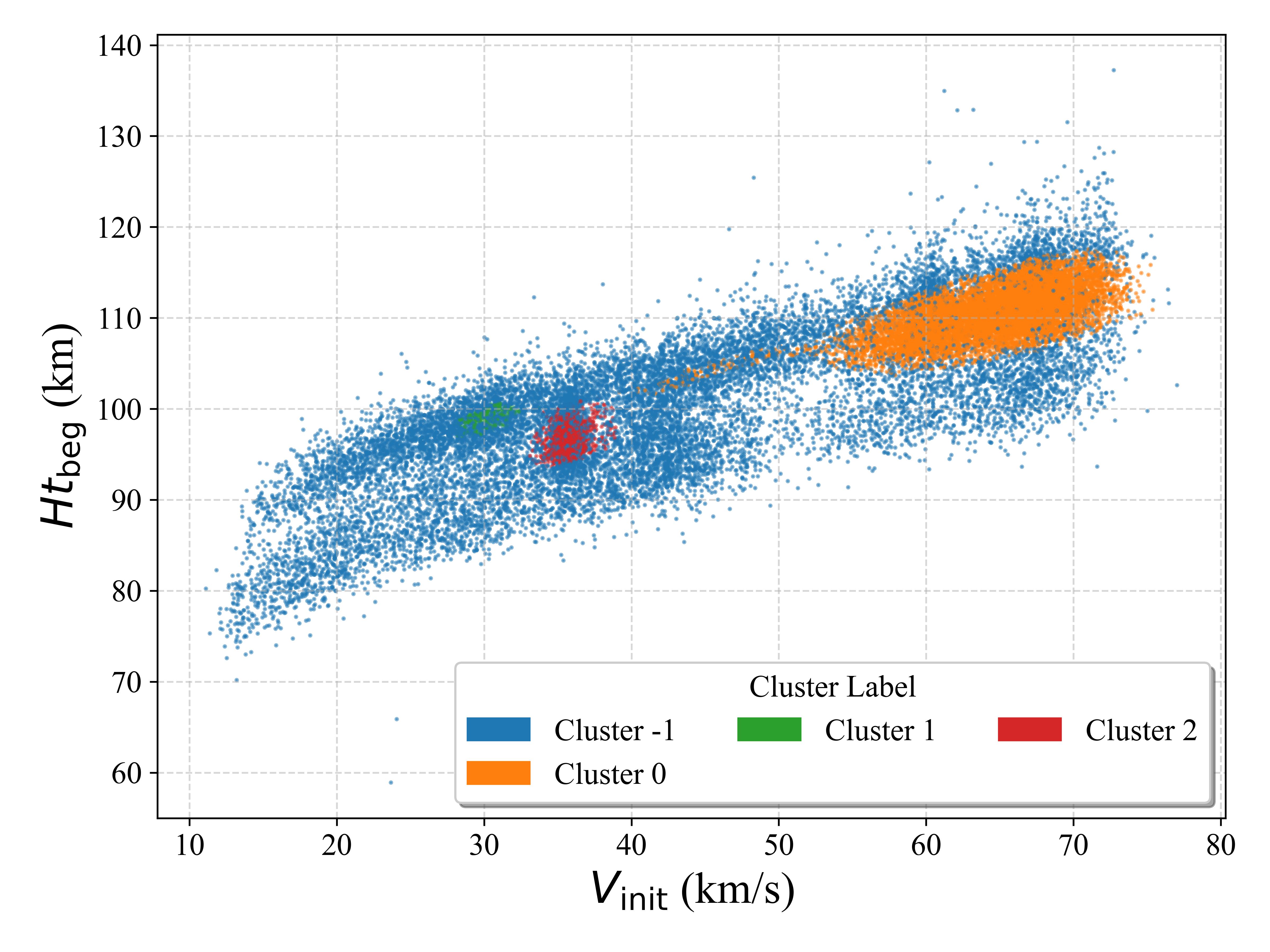}
        \caption{\centering FA + \texttt{min\_cluster\_size} = 100\\$S=-0.03$}
        \label{hdbscan_b}
    \end{subfigure}
    \hfill
    \begin{subfigure}[t]{0.48\textwidth}
        \centering
            \includegraphics[width=\linewidth]{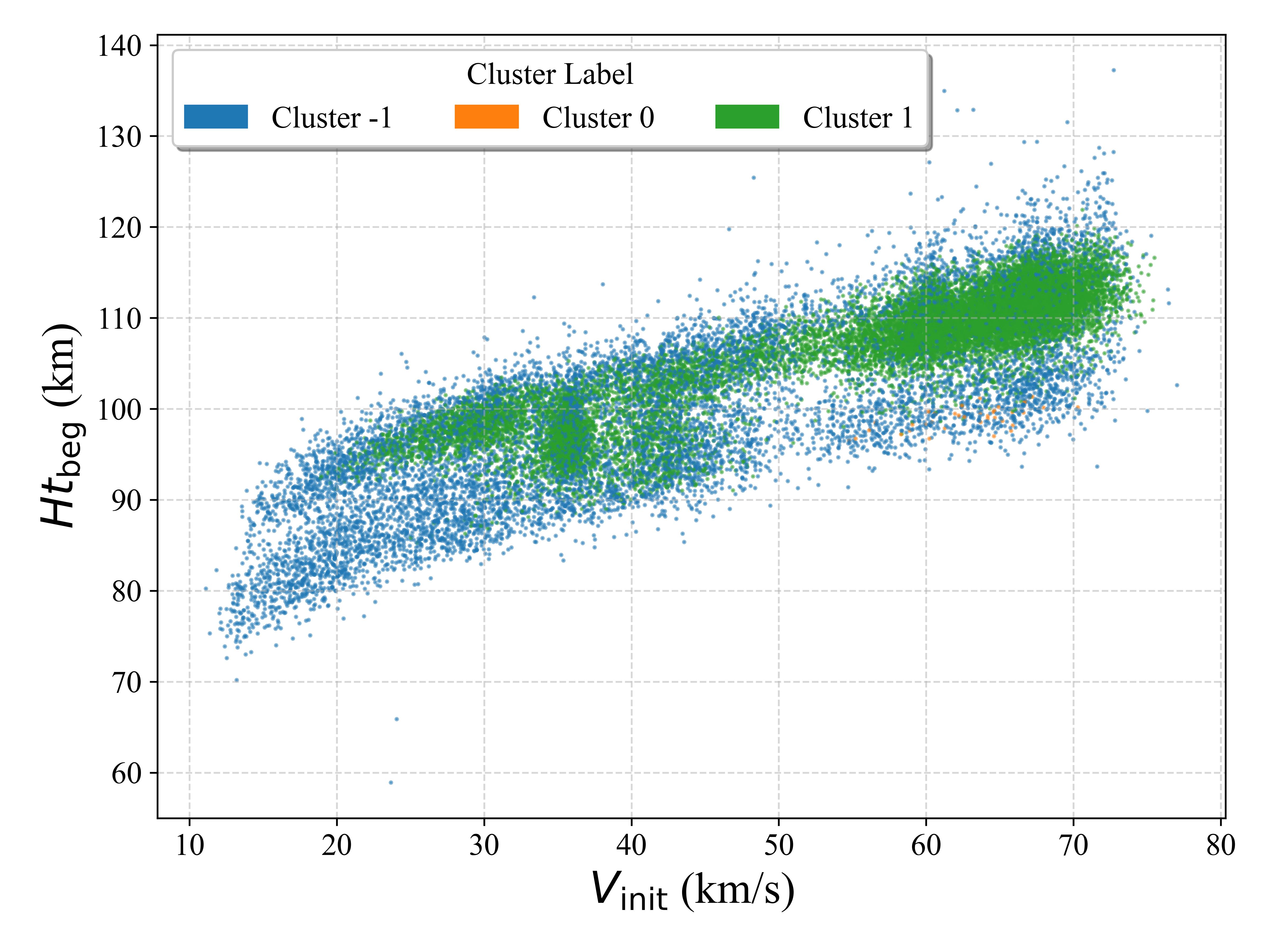}
        \caption{\centering PCA + \texttt{min\_cluster\_size} = 20\\$S=0.03$}
        \label{hdbscan_c}
    \end{subfigure}
    \hfill
    \begin{subfigure}[t]{0.48\textwidth}
        \centering
            \includegraphics[width=\linewidth]{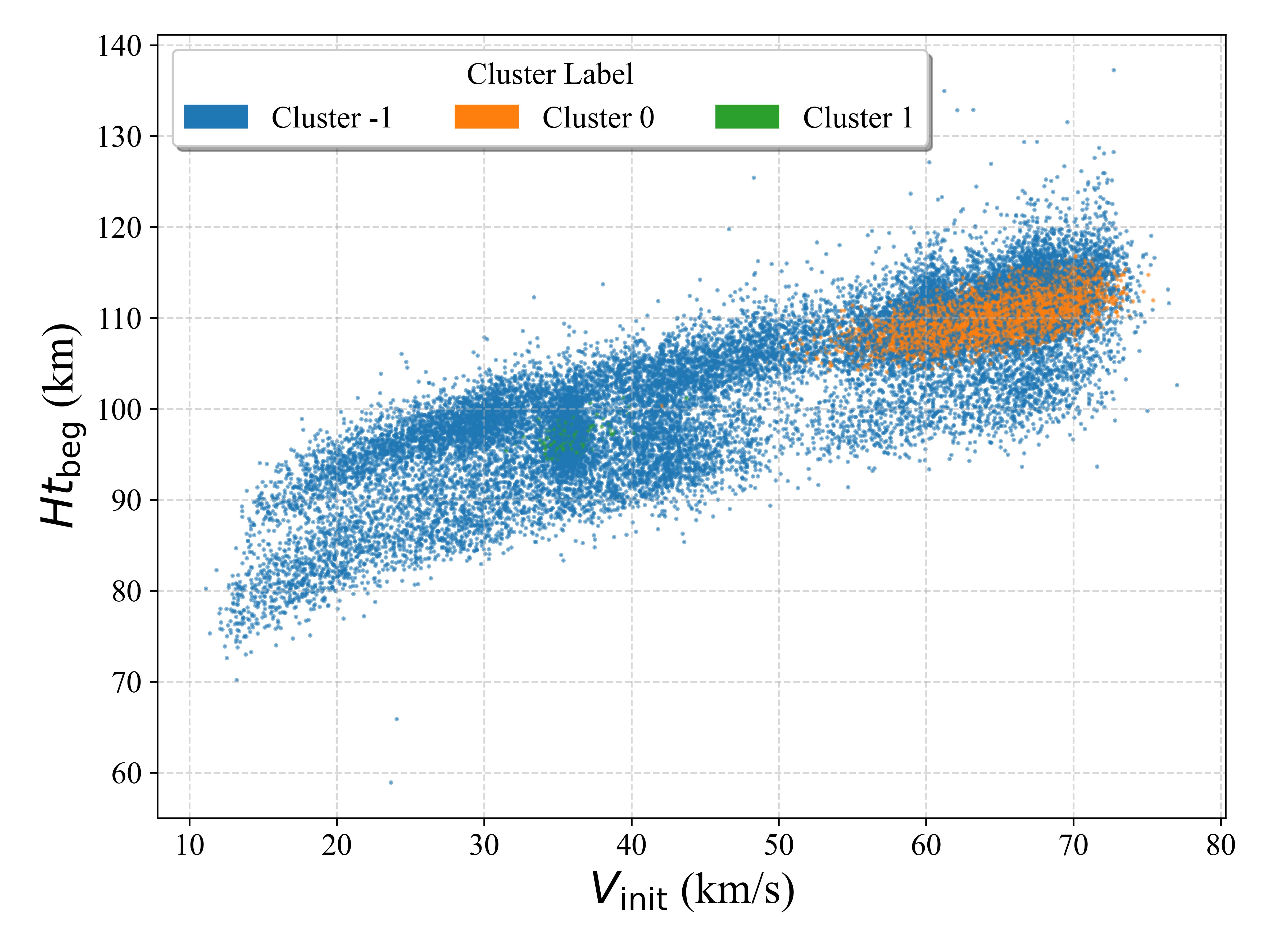}
        \caption{\centering PCA + \texttt{min\_cluster\_size} = 100\\$S=-0.16$}
        \label{hdbscan_d}
    \end{subfigure}
    \caption{HDBSCAN clustering results in $Ht_\mathrm{beg}$ vs. $V_\mathrm{init}$ space for different combinations of dimensionality reduction techniques and \texttt{min\_cluster\_size} values. $S$ indicates modest clustering performance across all configurations. Panels a–b show FA results and panels c–d show PCA results. Cluster number $-1$ is assigned by the algorithm to noise points. All other cluster numbers represent definitive clusters identified by HDBSCAN.}
    \label{fig:hdbscan}
\end{figure}

For both dimensionality reduction techniques, we varied the \texttt{min\_cluster\_size} hyper-parameter with values of 5, 10, 20, 25, 50, 75, 100, and 150 to assess its effect on clustering. When \texttt{min\_cluster\_size} $<$ 20, HDBSCAN produced one large cluster with several very small clusters and a broad region of noise points  (as in Figure~\ref{hdbscan_a}) or one large cluster (as in Figure~\ref{hdbscan_c}) with a broad region of noise points. We found that, for higher values of \texttt{min\_cluster\_size} ($\geq$ 20; Figure~\ref{fig:hdbscan}), HDBSCAN consistently identified regions corresponding to known meteor showers (as in Figure~\ref{hdbscan_b} and \ref{hdbscan_d}), including the Geminids and the combined region of the Perseids, Orionids, and Eta Aquariids.

However, when compared to the groupings produced by $K_b$ in Figure~\ref{kb_ht_vel}, the HDBSCAN clusters did not have any physical relevance: the vast majority of points were assigned as noise while many of the clusters it found had few members. This indicated that HDBSCAN did not reliably recover structure related to meteoroid strength in our data and was therefore excluded from further analysis. In the studies above where HDBSCAN performed well, a time-related parameter (e.g. solar longitude) was included. We suspect that the absence of such a parameter here may be a reason the method was not well suited to our approach and interest in meteor strength properties as opposed, for example, to the identification of annual showers.

\section{Results} \label{sec:results}

\begin{figure}
    \centering
        \includegraphics[width=\textwidth]{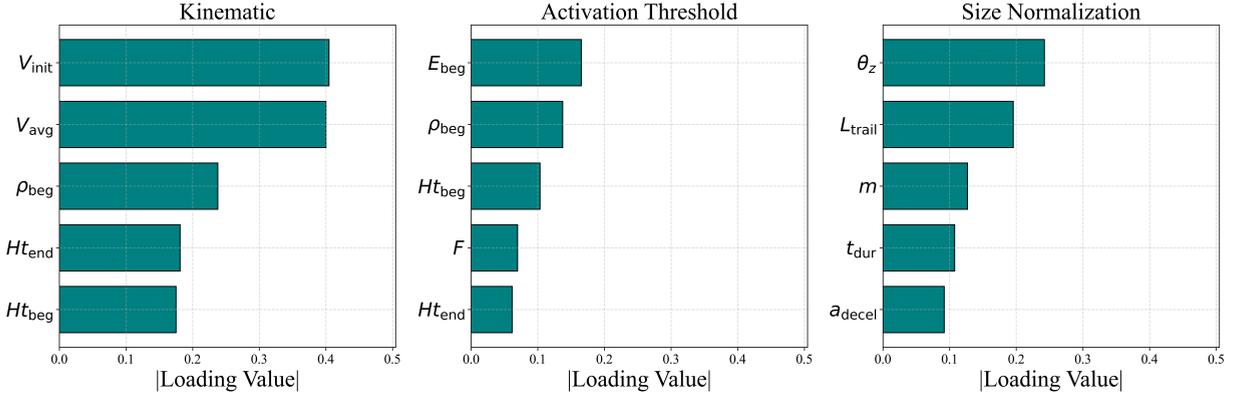}
    \caption{Magnitudes of features comprising the top five factor loadings for each of the kinematic, activation threshold, and size normalization factors.}
    \label{factor_bar}
\end{figure}

\subsection{Factor Analysis} \label{subsec:red_dimensions} 
3 factors produced the best defined and realistic clusters in our final model (Section~\ref{subsec:rej_dimReduc}). The latent variables represented by each factor were interpreted based on the dominant set of parameters present in each factor. Figure~\ref{factor_bar} visualizes the magnitudes of the top five features for each factor. These three factors are associated with meteors' kinematics, activation thresholds, and size normalization. The kinematic factor contains information on a meteor's motion through the atmosphere. Activation threshold quantifies a meteor's resistance to the onset of ablation and fragmentation. The size normalization factor contains indicators of a meteor's size and viewing geometry. 

For simplicity, we focus this section on the five features with the largest factor loadings for each factor. See Table~\ref{tab:fa_params} in Appendix \ref{model_params} for each feature's full factor coefficient. The kinematic factor loaded positively on $\rho_\mathrm{beg}$ (0.238) and negatively on $V_\mathrm{init}$ (-0.405), $V_\mathrm{avg}$ (-0.400), $Ht_\mathrm{end}$ (-0.181), and $Ht_\mathrm{beg}$ (-0.175). Activation threshold loaded positively on $E_\mathrm{beg}$ (0.165) and $\rho_\mathrm{beg}$ (0.137), and negatively on $Ht_\mathrm{beg}$ (-0.103), $F$ (-0.070), and $Ht_\mathrm{end}$ (-0.062). Size normalization loaded positively on $L_\mathrm{trail}$ (0.196), $m$ (0.127), and $t_\mathrm{dur}$ (0.108), and negatively on $\theta_z$ (-0.242) and $a_\mathrm{decel}$ (-0.092). Three features, $M_\mathrm{abs}$, $F$, and $a_\mathrm{decel}$, have factor loadings $<$ $\left| 0.100 \right|$ across all 3 factors which indicates they may not have a significant effect on factor structure. We analyze the effect of removing these three features from the feature set in the next section (Section~\ref{subsubsec:dependencies}).

\subsubsection{Feature Dependencies and Their Impact on FA} \label{subsubsec:dependencies}

\begin{figure}
    \centering
        \includegraphics[width=0.8\textwidth]{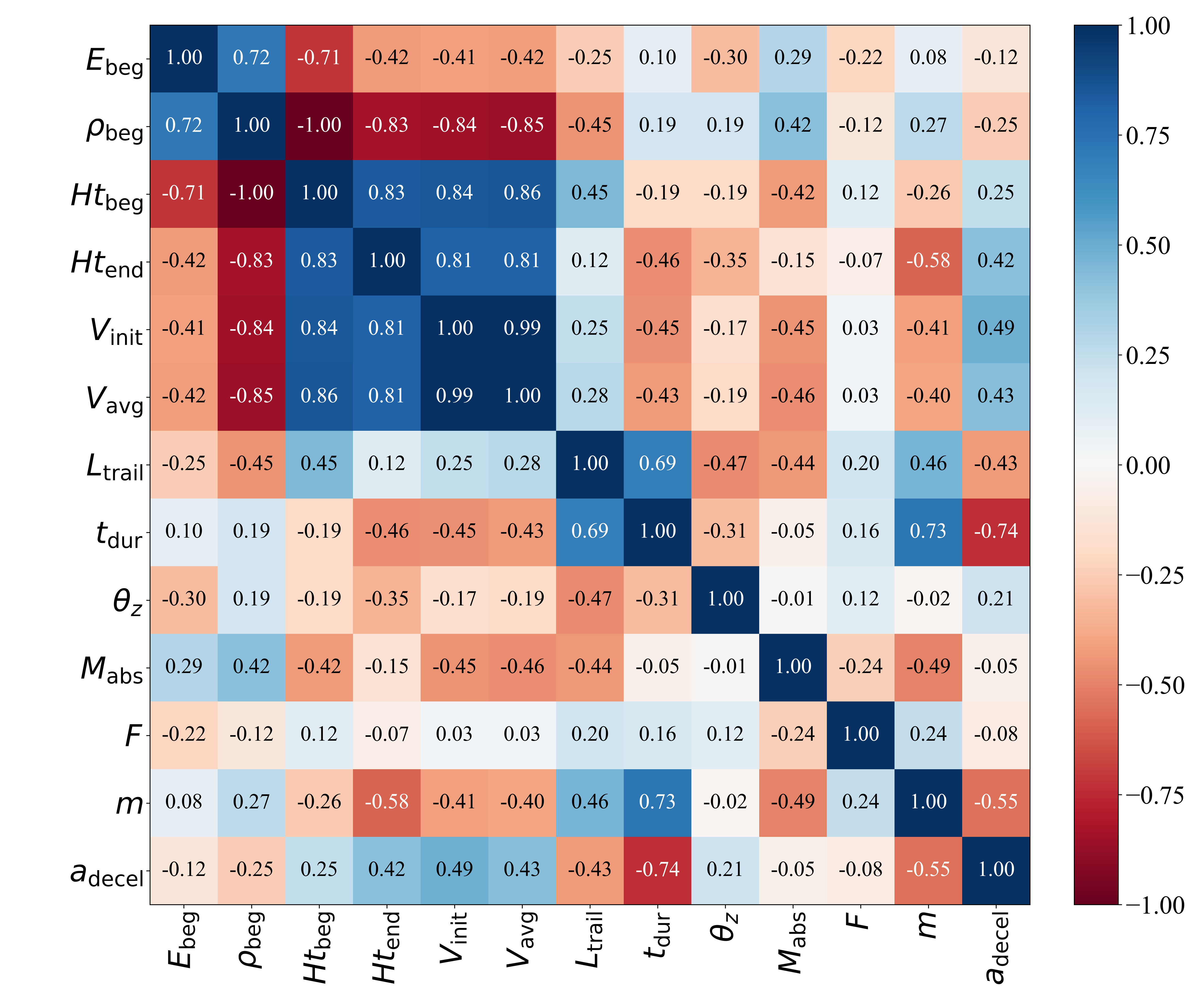}
    \caption{Correlation matrix of the 13 features described in Table~\ref{features}, highlighting structured correlations that underlie the three latent factors identified through FA: Kinematics, Activation Threshold, and Size Normalization.}
    \label{corr_matrix}
\end{figure}

Several of the features we included are not independent. Specifically, $E_\mathrm{beg}$, $\rho_\mathrm{beg}$, $L_\mathrm{trail}$, and $a_\mathrm{decel}$ are derived quantities that depend on other observable parameters. $E_\mathrm{beg}$ is computed by first integrating atmospheric density from a meteor's entry into the atmosphere to the height at which erosion begins, therefore depending on $Ht_\mathrm{beg}$. This atmospheric density integral is then used in conjunction with $V_\mathrm{init}$ and $\theta_z$ to compute a meteor's kinetic energy per unit cross section. $\rho_\mathrm{beg}$ is derived from $Ht_\mathrm{beg}$ through an atmospheric model while $a_\mathrm{decel}$ and $L_\mathrm{trail}$ are both functions of $V_\mathrm{init}$ and/or $V_\mathrm{avg}$ together with $t_\mathrm{dur}$. FA explicitly models shared covariance which is likely why it worked well in this case (see Section~\ref{subsec:fa_discussion}), however strongly correlated or derived features influence the structure of the covariance matrix and affect the resulting factor loadings. Figure~\ref{corr_matrix} shows a correlation matrix for all 13 features we used (Table~\ref{features}).

One method used to compare FA results is Tucker's Congruence Coefficient \citep{Tucker1951}, which is mathematically equivalent to the cosine similarity between factor loading vectors. Each element of the congruence matrix ($\phi_{ij}$) is computed as the dot product between the loading vectors of two factors, normalized by the product of their vector magnitudes. This normalization removes any dependence on the magnitude of the loading vectors so that only their relative orientation matters. Mathematically, this is:

\begin{equation}
    \phi_{ij} =
\frac{\mathbf{W}_{i} \cdot \mathbf{Y}_{j}}
{\|\mathbf{W}_{i}\| \, \|\mathbf{Y}_{j}\|}
\label{equ:tucker}
\end{equation}

Where $\mathbf{W}$ is the factor loading matrix derived from the full feature set and $\mathbf{Y}$ represents the factor loading matrix derived from a reduced feature set. The comparison is restricted to common variables, such that only the features present in $\mathbf{Y}$ are used from $\mathbf{W}$ when computing $\phi_{ij}$. This allows the similarity of the resulting factor structures to be evaluated after removing selected features

$\phi_{ij} \geq$ 0.95 indicates factors that are essentially equal or identical, 0.85 $\leq \phi_{ij} <$ 0.95 corresponds to factors with fair similarity, and $\phi_{ij} <$ 0.85 indicates there is little to no similarity between the two factors \citep{lorenzo2006}. We computed $\phi$ for three sets of reduced features. Set A removed several features dependent upon other observables from Table~\ref{features}: $E_\mathrm{beg}$, $\rho_\mathrm{beg}$, $L_\mathrm{trail}$, and $a_\mathrm{decel}$. Set B removed the two features not dependent on height or velocity terms: $L_\mathrm{trail}$, and $a_\mathrm{decel}$. Set C removed the three features with factor loadings $<$ $\left| 0.100 \right|$ across all 3 factors: $M_\mathrm{abs}$, $F$, and $a_\mathrm{decel}$. Tucker congruence matrices comparing factor loadings from the full feature set to those from sets A, B, and C are shown in Table~\ref{tab:congruence}. Factors were matched across models by identifying factors that are essentially equal or fairly similar to the full model factors.

The results from Sets A and B highlight the importance of $E_\mathrm{beg}$ and $\rho_\mathrm{beg}$ in defining the activation threshold factor. In Set A, where both variables are removed, activation threshold is not recovered. When these features are reintroduced in Set B, the factor is nearly fully restored (Set B $\phi_{2,3}=0.975$). Although $E_\mathrm{beg}$ and $\rho_\mathrm{beg}$ are strongly correlated with $Ht_\mathrm{beg}$ (Figure~\ref{corr_matrix}), the inability of $Ht_\mathrm{beg}$ alone to reproduce the factor indicates that these derived quantities capture additional information about the meteoroid's interaction with the atmosphere.

Kinematics remained highly stable across all three reduced feature sets, demonstrating that it is robust to the removal of $\rho_\mathrm{beg}$. Size normalization shows moderate similarity to Factor 3 in Set A (Set A $\phi_{3,3}=0.864$) and is identical to Factor 2 in Set C (Set C $\phi_{3,2}=1.000$). However, this factor is not recovered in Set B, indicating that $L_\mathrm{trail}$ and $a_\mathrm{decel}$ contribute significantly to the structure captured by this factor.

Although the factor structure produced by Set C is nearly identical to that of the full model, we retain $M_\mathrm{abs}$, $F$, and $a_\mathrm{decel}$ because the overall clustering score is slightly higher when these variables are included ($S=0.292$) than when they are removed ($S=0.290$). The similarity of these results suggest that future work could reasonably omit these variables without substantially altering the factor structure and clustering results.

A primary limitation of this method is that it only focuses on their directional similarity and does not capture differences in loading magnitudes. To address this, we computed the mean difference between factors deemed essentially equal  ($\phi_{ij} \geq$ 0.95). The mean difference across all six comparisons was 0.003, indicating that factors identified as congruent also displayed very similar factor loading magnitudes.

\begin{table*}[t]
\centering
\caption{Tucker congruence matrices comparing factor loadings derived from the full feature set (Table~\ref{features} and \ref{tab:fa_params} in Appendix~\ref{model_params}) and three reduced feature sets (Section~\ref{subsubsec:dependencies}). Rows represent the factors for the full model (all 13 features) and columns represent factors for the reduced feature sets. The removed line indicates which features were removed from the full feature set to create the reduced feature set. Each entry ($\phi_{ij}$) gives the Tucker congruence coefficient between factor $i$ of the full feature model and factor $j$ of the reduced feature model. F1, F2, and F3 denote Factors 1--3 for each reduced feature set. Factors that are fairly similar or essentially equal are bolded (see Section~\ref{subsubsec:dependencies}).}
\label{tab:congruence}

\begin{tabular}{lccc ccc ccc}
\toprule
 & \multicolumn{3}{c}{Set A} & \multicolumn{3}{c}{Set B} & \multicolumn{3}{c}{Set C} \\
\cmidrule(lr){2-4}\cmidrule(lr){5-7}\cmidrule(lr){8-10}
 & \multicolumn{3}{c}{\scriptsize Removed: $E_\mathrm{beg}$, $\rho_\mathrm{beg}$, $L_\mathrm{trail}$, and $a_\mathrm{decel}$}
 & \multicolumn{3}{c}{\scriptsize Removed: $L_{\mathrm{trail}}, a_{\mathrm{decel}}$}
 & \multicolumn{3}{c}{\scriptsize Removed: $M_\mathrm{abs}$, $F$, and $a_\mathrm{decel}$} \\
Full Model Factor & F1 & F2 & F3 & F1 & F2 & F3 & F1 & F2 & F3 \\
\midrule

Kinematics &
\textbf{0.999} & 0.140 & -0.184 &
\textbf{1.000} & 0.099 & 0.460 &
\textbf{1.000} & -0.095 & 0.495 \\

Activation Threshold &
0.327 & -0.177 & -0.164 &
0.428 & -0.136 & \textbf{0.975} &
0.444 & -0.178 & \textbf{0.999} \\

Size Normalization &
-0.041 & 0.578 & \textbf{0.864} &
-0.086 & 0.701 & -0.279 &
-0.081 & \textbf{1.000} & -0.172 \\

\bottomrule
\end{tabular}
\end{table*}

\subsection{Clustering} \label{subsec:clustering_results}

\begin{figure}[!htbp]
    \centering
    % --- (a) 3 Clusters ---
    \begin{subfigure}[t]{0.78\linewidth}
        \centering
            \includegraphics[width=\linewidth]{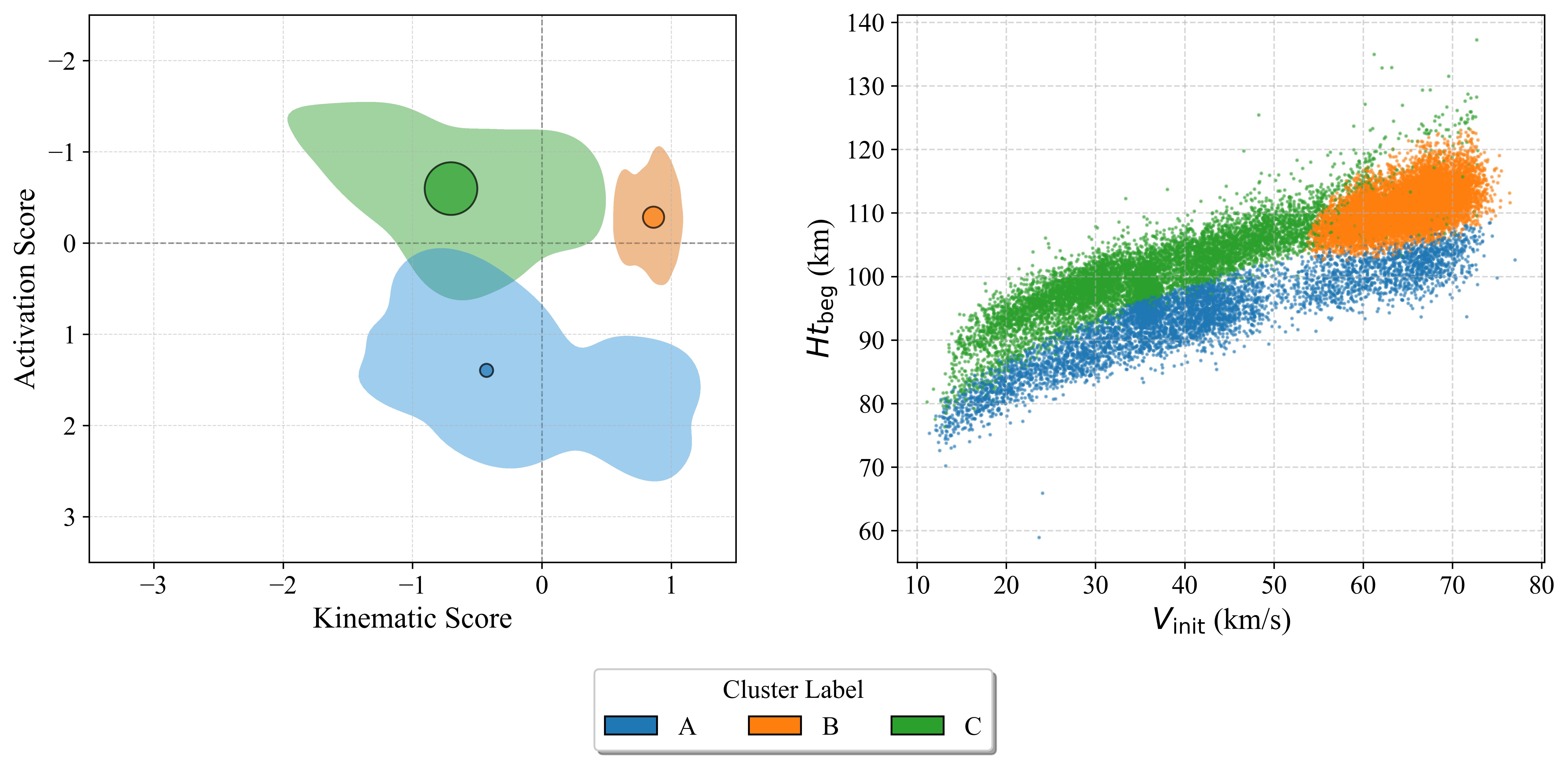}
        \caption{3 Clusters ($S=0.29$)}
        \label{3_cluster_result}
    \end{subfigure}

    % --- (b) 6 Clusters ---
    \begin{subfigure}[t]{0.78\linewidth}
        \centering
            \includegraphics[width=\linewidth]{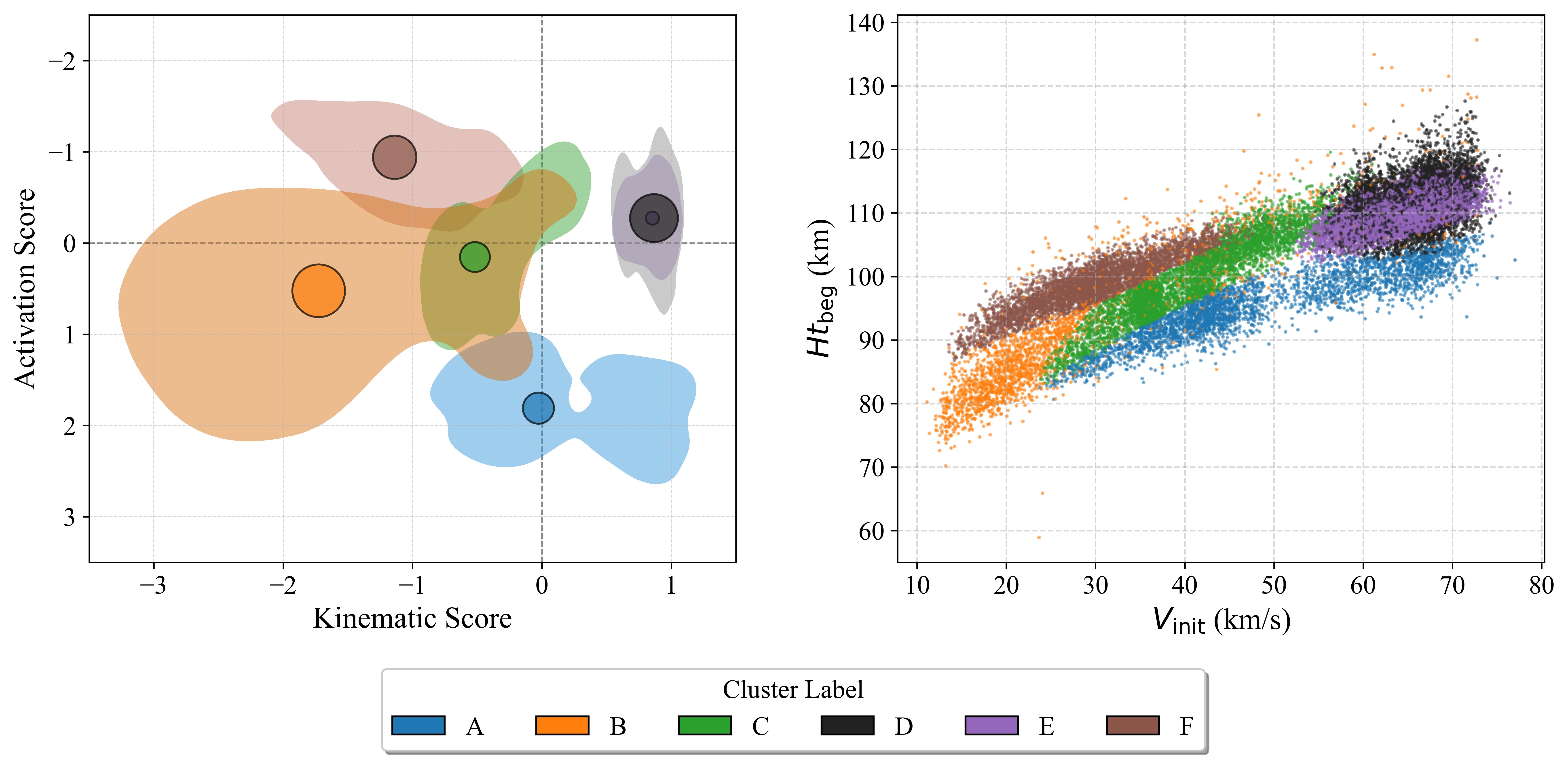}
        \caption{6 Clusters ($S=0.24$)}
        \label{6_cluster_result}
    \end{subfigure}

    % --- (c) 11 Clusters ---
    \begin{subfigure}[t]{0.78\linewidth}
        \centering
            \includegraphics[width=\linewidth]{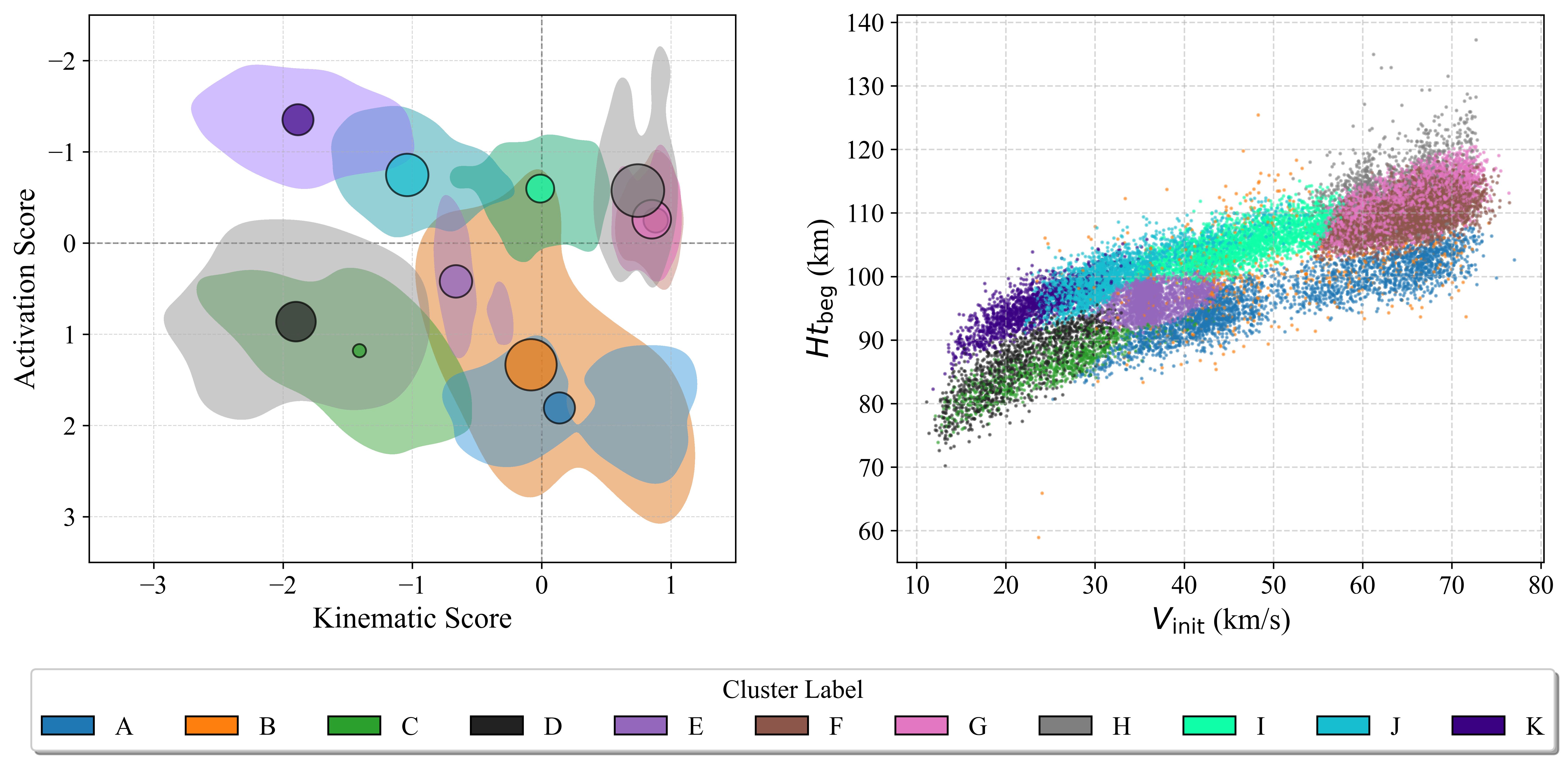}
        \caption{11 Clusters ($S=0.23$)}
        \label{11_cluster_result}
    \end{subfigure}
    \caption{Results from applying a GMM to FA reduced LO-CAMS data for (a) 3 clusters, (b) 6 clusters, and (c) 11 clusters. The left panel shows median activation scores (y-axis) vs. median kinematic scores (x-axis) for each cluster. The size of the scatter points scale with the size factor (smaller circle = smaller median size factor). The contours contain 66.7\% (1$\sigma$) of the data points for each cluster. The right panel shows $Ht_\mathrm{beg}$ vs. $V_\mathrm{init}$ scatter plots.}
    \label{fig:cluster_subfigs}
\end{figure}

We applied 3, 6, and 11 cluster GMMs to the FA reduced dataset (Section~\ref{subsec:rej_clustering}). The left panels of Figure~\ref{fig:cluster_subfigs} show the median factor scores for each cluster. Median factor scores by cluster for each of the three models are summarized in Tables~\ref{tab:3_cluster_medians},~\ref{tab:6_cluster_medians}, and~\ref{tab:11_cluster_medians} in Appendix~\ref{cluster_medians}. Because the velocity parameters load inversely on factor scores, we reversed the sign of the kinematic factor such that positive values correspond to faster objects. The sizes of the points represent median values of the size normalization factor for each cluster. The contours show the range of the activation and kinematic factors and encapsulate 66.7\% ($1\sigma$) of cluster data points. We also inverted the activation axis so that the cluster ordering roughly matches that seen in the $Ht_\mathrm{beg}$ vs. $V_\mathrm{init}$ plots from left to right. The right panel shows the corresponding  $Ht_\mathrm{beg}$ vs. $V_\mathrm{init}$ visualizations for each model. Across all three models, we found that the activation factor emerged as the dominant parameter governing inferred meteor composition, while the kinematic and size factors act primarily as normalization terms. As seen in the left panel of Figure~\ref{fig:cluster_subfigs}, asteroidal/carbonaceous groups in terms of $K_{b}$ generally exhibit positive activation scores while cometary groups trend negative. Cluster labels are therefore ordered in terms of median activation score, from highest to lowest, which is indicative of the material hardness of constituent meteors. 

The 3 cluster model (Figure~\ref{3_cluster_result}) showed three clearly separated groups that correspond to high and low velocity cometary meteors, and the carbonaceous population identified by $K_b$ in Figure~\ref{kb_ht_vel}. The 6 cluster model (Figure~\ref{6_cluster_result}) further subdivided these groupings and began to provide a more detailed view of the relationships among the kinematic, activation threshold, and size normalization factors. The 11 cluster model (Figure~\ref{11_cluster_result}) captured nuanced distinctions among the three $K_b$ regimes. Interpretation was slightly complicated by clusters that displayed overlapping or transitional characteristics. Nonetheless, many of these clusters aligned with physically realistic groupings inferred from $K_{b}$ and meteor showers. This suggests that increasing the number of clusters captures physically meaningful subtleties that are not resolved in simpler models. Going forward, we adopt the 11 cluster model to provide specific insights on meteor properties.

Within the 11 cluster model, clusters B and F exhibit behavior that does not translate cleanly into physically meaningful populations (Figure~\ref{11_cluster_result}). In factor space (left panel), cluster B spans a broad region, indicating a wide range of factor scores and suggesting it represents a heterogeneous grouping dominated by larger meteoroids. These meteoroids span a wide range of heights and velocities in $Ht_\mathrm{beg}$ vs. $V_\mathrm{init}$ space (right panel) and do not occupy a well-defined region as the other clusters do. Cluster F occupies nearly the same region as cluster G in both panels and therefore does not appear to be physically distinct. 

To address these ambiguities, meteoroids initially assigned to clusters B and F are reassigned to their next most probable cluster based on the posterior probabilities returned by the GMM. The rationale behind simply reassigning clusters based on posterior probabilities is motivated by the fact that the GMM posteriors already encode the multivariate structure of the data, incorporating the combined influence of all observables through the learned covariance of each component. The secondary posterior probability represents the most statistically consistent alternative classification once the original class is removed. An alternative method could be to penalize assignments that are inconsistent with a meteoroid’s activation factor score, for example by weighting each posterior probability by the object’s distance from the median activation value of each cluster. Such a scheme would reinforce the observed importance of activation scores on cluster assignment. However, it introduces an additional tunable parameter and imposes a model-dependent prior on the clustering outcome. We therefore defer this extension to future work and adopt the posterior-based reassignment as the most conservative and reproducible choice.

With the merging of events in clusters B and F into their next most probable clusters, the 11 cluster solution is reduced to nine physically interpretable classes ordered by inferred material hardness. This classification scheme, hereafter denoted as $H_{\mathrm{class}}$, serves as a data-driven extension of $K_{b}$. In Section~\ref{sec:discussion}, we apply $H_{\mathrm{class}}$ to well characterized meteor showers, analyze the orbital properties of each class, and use these results to interpret the material composition of each $H_{\mathrm{class}}$ group. These same methods are then applied to a 9 cluster FA–GMM model to evaluate if adopting a 9 cluster model from the outset would have been effective.

\FloatBarrier

\FloatBarrier

\section{Discussion} \label{sec:discussion}

\subsection{Application to Meteor Showers} \label{subsec:showers}

We applied $H_{\mathrm{class}}$ to nine well-studied meteor showers contained in the GMN dataset (Section~\ref{subsec:data}) to evaluate how it classifies meteors with inferred densities and understand how it assigns shower members to our newly constructed classifications. This application also serves to validate our interpretation of the three latent factors and aid in the physical interpretation of the resulting classification scheme. The showers included were the Perseids (PER; 72,094 events), Orionids (ORI; 30,841), Eta Aquariids (ETA; 19,480), Lyrids (LYR; 6,275), Tau Herculids (TAH; 1,115), Geminids (GEM; 47,449), Quadrantids (QUA; 8,841), and Southern Delta Aquariids (SDA; 16,586), along with Iron candidates (19,991). Iron candidates were separated from the entire dataset using the criterion in \cite{Mills_2021}. Although the Iron candidates are not a meteor shower, we include them here for comparison as a physically distinct population (e.g. high density, low $Ht_{beg}$).

We did not apply the same quality filtering used in our training dataset, but we again removed all events containing ‘NaN’ and physically unrealistic values. 222,672 total events were associated with the nine showers listed above. In this application, we took advantage of the function's ability to output the probability that a given event falls into a particular cluster. We only considered shower events that had $\geq$ 80\% probability of cluster assignment and refer to these as confidently assigned meteors. We define retention as the percentage or number of meteors in each shower that were confidently assigned to a cluster.

The 80\% confidence threshold was chosen to ensure that only meteors with a high likelihood of belonging to their assigned cluster were included, allowing us to analyze shower events with minimal ambiguity in cluster membership. The drawback to this approach is that increasing the probability threshold noticeably decreases the number of meteors with confident assignments. A 90\% threshold retained only 18.6\% of shower meteors, whereas an 80\% threshold retained 45.4\%. Thus, we adopted an 80\% threshold as an effective balance between retention and confident assignment. 

\begin{figure}
    \centering
        \includegraphics[width=\linewidth]{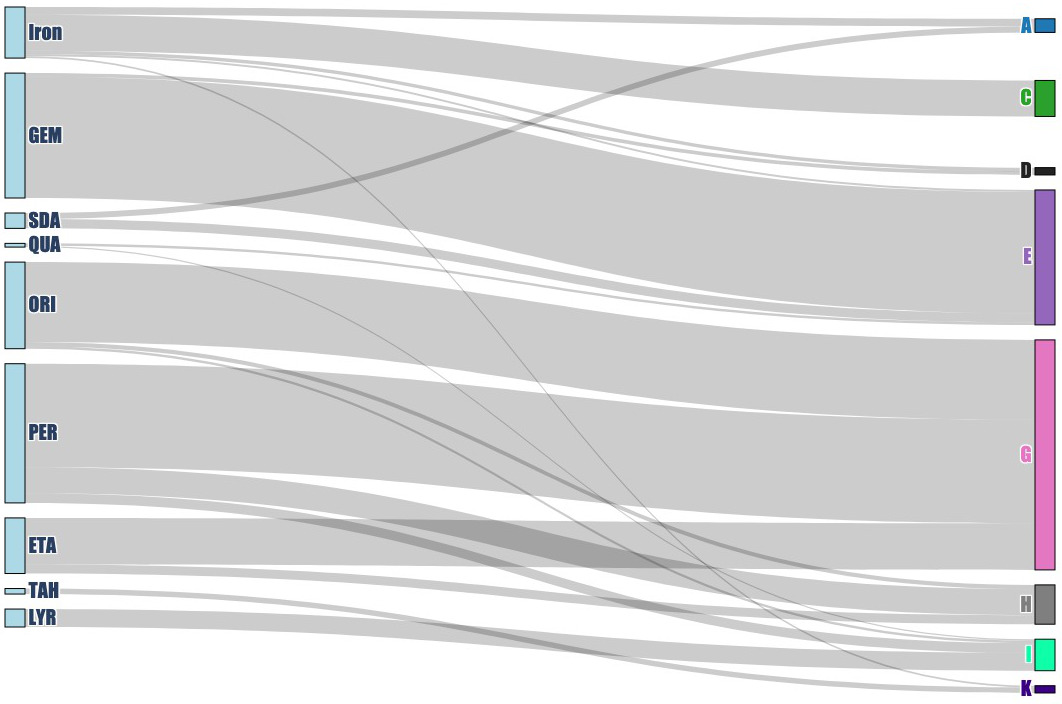}
    \caption{Sankey diagram of how the $H_{\mathrm{class}}$ model grouped the populations of 9 meteor showers in the full GMN dataset from 2018 December 10 through 2025 May 16. To improve the clarity of this figure, only flow lines with a minimum of 250 meteors are shown.}
    \label{fig:sankeydiagram}
\end{figure}

Figure~\ref{fig:sankeydiagram} shows a Sankey diagram illustrating how each shower was divided among the $H_{\mathrm{class}}$ groups. A Sankey diagram is a flow-style visualization that illustrates how quantities move or transition from one state to another. They are effective at showing how elements redistribute between groups or classifications. These diagrams are named after Captain Matthew Henry Phineas Riall Sankey who used this type of visualization to show the energy efficiency of a steam engine \citep{Sankey_1898}. The left side shows the nine meteor showers, where the heights of the light blue bars represent the proportion of each shower's meteors relative to the total population. On the right side is the $H_{\mathrm{class}}$ assigned by the model. The heights of the colored bars are proportional to the number of meteors in each class relative to the total population. The widths of the gray flow lines going from showers on the left to classes on the right are representative of the volume of meteors in each shower assigned to a group. Only flow lines with a minimum of 250 meteors are shown. There may be specific meteors that got misclassified relative to assignments from the GMN pipeline. Therefore, we focus on the overall groupings produced by the clustering algorithm. Tables~\ref{tab:cluster_counts_11} and~\ref{tab:cluster_percentages_11} in Appendix~\ref{sankey_stats} provide a full summary of shower meteor counts and cluster assignments used here.

The ORI, ETA, and PER each had $>$92\% of confidently assigned meteors distributed between two different $H_{\mathrm{class}}$ groupings: G, and H. ORI had the largest proportion of their meteors assigned to class G (91.8\%). ETA and PER respectively had 82.8\% and 74.2\% of their meteors assigned to class G.  In factor space, Class G meteors have moderate size scores, high kinematic scores, and negative activation. These scores map directly to defining physical behaviors of these events: high beginning heights, fast velocities, short trail lengths, and ablation in low density portions of the atmosphere. Class H represents a subclass of meteors populated by only small proportions of these showers: ORI (4.9\%), ETA (16.0\%), and PER (18.7\%). This classification seems to capture a subset of massive meteoroids in these showers. The similar $H_{\mathrm{class}}$ assignments of these showers are consistent with the fact that 1P/Halley is the parent of the ORI and ETA \citep{Whipple_1951, McIntosh_Hajduk_1983, McIntosh_Jones_1988, Egal_2020, Jenniskens2023} along with the measured cometary densities of all three showers \citep{Buccongello_2024, Vida_2024, BellotRubio_2002, Verniani_1967}.

99.8\% of confidently assigned TAH meteors were placed into $H_{\mathrm{class}} =$ K. This shower showed 98.7\% retention after applying the 80\% cutoff for confident cluster assignment, nearly all of which were associated with the 2022 outburst \citet{egal2023modeling}. $H_{\mathrm{class}} =$ K meteors exhibit moderate size scores and amongst the lowest activation and kinematic scores of all nine classes. Physically, this class corresponds to bright, slow meteors with relatively high beginning and ending heights, indicative of very low bulk densities and very fresh cometary material. These characteristics are consistent with modeling of the TAH by \citet{egal2023modeling}. Despite only moderate size factor scores, $H_{\mathrm{class}} =$ K also ranks in the upper quartile for mass, suggesting meteoroids in this group tend to be relatively large. This interpretation agrees with \citet{Ye_2022}, who found that the brightest meteors observed during the 2022 TAH outburst were likely produced by centimeter-scale or larger particles.

The Iron group was dominated by $H_{\mathrm{class}} =$ C, with 70.4\% of confidently assigned events assigned to this class. Characteristics of this class are representative of dense meteors that penetrate deeply into the atmosphere and exhibit high activation. An additional 14.9\% of Irons were assigned to $H_{\mathrm{class}} =$ A. Class A has the highest activation scores, corresponding to dense, carbonaceous type meteoroids. This finding might indicate that the simple criterion used for classifying iron meteoroids from \citet{Mills_2021} includes some non-iron objects.

The LYR had 92.6\% of confidently assigned meteors grouped into $H_{\mathrm{class}} =$ I, isolating them into a separate group relative to all other showers. This may indicate that they are compositionally unique, supported by the fact that they are the strongest shower produced by an LPC \citep{lyytinen2003meteor}. This class of meteors exhibits a cometary activation score with moderate size and kinematics.

The GEM had 96.3\% of meteors assigned to $H_{\mathrm{class}}=$ E. This class also has some contributions from the SDA and QUA showers, indicative of their similar carbonaceous properties. However, the class was comprised of $\sim$90\% GEM meteors. $H_{\mathrm{class}} =$ E exhibits a positive activation (asteroidal) along with moderate size and kinematics consistent with ablation at high atmospheric densities and low beginning heights. These characteristics are indicative of carbonaceous type meteoroids and align with measured bulk densities \citep{Buccongello_2024, Babadzhanov_2002, BellotRubio_2002, Verniani_1967}.

The SDA and QUA showers both showed the lowest retention rates (20.8\% and 13.4\%, respectively) among all of the showers. The low assignment probabilities can likely be attributed to their broad range of physical properties and complex dynamical evolution. Comet 96P/Machholz was first proposed as related to the QUA by \cite{McIntosh_1990}, who suggested that the comet's past orbits were similar to the shower's current mean orbit. \cite{Jenniskens_2004} later proposed asteroid 2003 EH1 as a parent body of the QUA, based on its present-day orbital similarity to the shower. Numerical integrations of the asteroid conducted in \cite{Babadzhanov_2008} investigated whether meteoroids ejected from this object can intersect Earth at different times, other than in January when the annual QUA meteor shower occurs. Their results showed that meteoroids from 2003 EH1 intersect Earth at eight different arguments of perihelion, corresponding to eight distinct showers. Most of these predicted showers were associated with observed streams, including the SDAs. Because comet 96P/Machholz and 2003 EH1 occupy similar regions of orbital phase space, \cite{Kanuchova_2007} and \cite{Neslusan_2013a} both suggested that 96P may also be a parent of the QUA stream and, by extension, the SDAs. \cite{Neslusan_2013} further showed that the 96P complex has four overlapping filaments associated with the 2003 EH1 complex, two of which correspond to the QUA and SDA showers. That study also found that the perihelion distance of the 96P/2003 EH1 shower complex can vary from 0.1 AU to 1.0 AU within a single libration cycle. These results present two possible explanations for the wide range of physical properties observed in the SDA and QUA streams. One is that both streams contain particles originating from two different parent bodies: one cometary and the other asteroidal. The second is that the large variation in perihelion distance subjects meteoroids to differing levels of thermal processing, producing a broad range of physical characteristics within the streams. The largest proportion of each shower were grouped into $H_{\mathrm{class}} =$ E showing similar characteristics to the GEM. 54.1\% of SDA and 38.7\% of QUA were assigned to this class. Most of the remaining SDA events (34.6\% of the total) were assigned to $H_{\mathrm{class}} =$ A, reflecting a high activation for a portion of this shower. The QUA had the remaining 61.3\% of confidently assigned meteors assigned to four $H_{\mathrm{class}}$ groupings: A (9.2\%), D (12\%), I (24\%), and J (16.1\%).

\begin{figure}
    \centering
        \includegraphics[width=\linewidth]{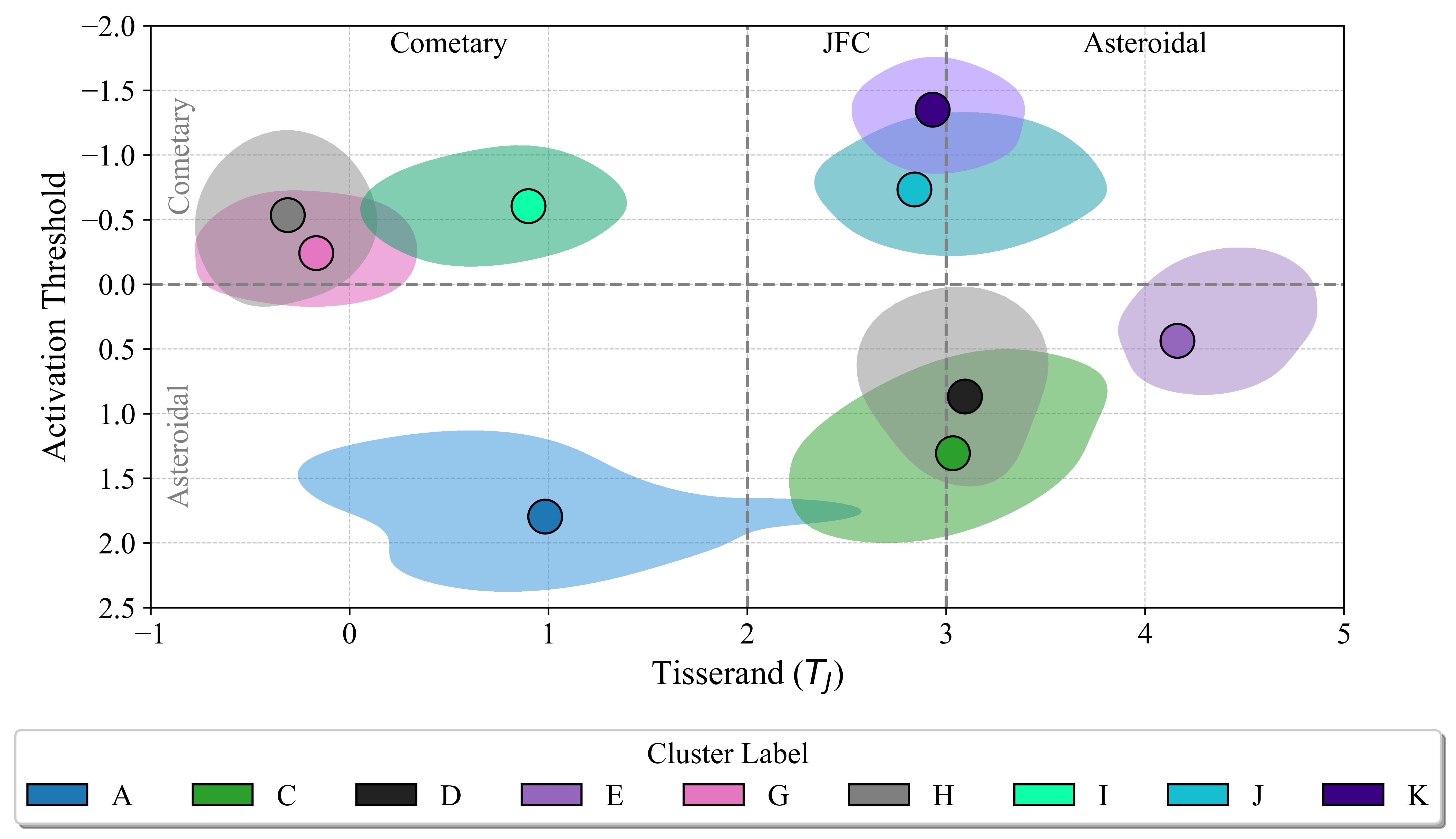}
    \caption{Evaluation of $H_{\mathrm{class}}$ clusters in orbital space for the LO-CAMS 2023 dataset. The median Tisserand parameter \citep{Levison_1996} is plotted along the x-axis, where the annotations along the top of this axis indicate orbital classifications. The y-axis shows activation threshold scores, where annotations correspond to inferred material hardness (negative = Cometary, positive = Asteroidal). Colors denote cluster assignments and shaded contours enclose the central 33\% of events in each cluster.}
    \label{fig:tissJplots}
\end{figure}

\subsection{Clusters in Orbital Space} \label{subsec:orbElements}

The Tisserand parameter with respect to Jupiter ($T_{J}$) incorporates three fundamental orbital elements ($a$, $e$, and $i$) and categorizes bodies into dynamical classes. Objects with $T_{J} > 3$ have orbits that are entirely interior to Jupiter and are considered asteroidal. Those with $2 < T_{J} < 3$ are on Jupiter Family Comet (JFC) type orbits while $T_{J} < 2$ represents objects with longer orbital periods \citep{Levison_1996}. These dynamical groupings provide a meaningful basis for evaluating assignments from $H_{\mathrm{class}}$. Figure~\ref{fig:tissJplots} shows the LO-CAMS 2023 dataset with median Tisserand parameter for each cluster on the x-axis and median activation score on the y-axis. The contours enclose the highest-density regions of this parameter space, corresponding to the 33.3\% density levels of the underlying event distribution. The plots are annotated with orbital classes described along the top of the x-axis and the structural groupings inferred by activation score on the y-axis. Events with asteroidal material properties occupy the region of activation $> 0$, whereas the cometary regime is $< 0$. This delineation is a convenient and physically relevant outcome that naturally emerged from FA, thus lending confidence to the diagnostic utility of our methodology.

Overall, cometary and asteroidal separations are visible while individual clusters occupy meaningful regions within both the activation factor and $T_{J}$. $H_{\mathrm{class}}=$ A appears to represent asteroidal material on cometary orbits. In the shower analysis, its population was dominated by the SDA with contributions from the QUA as well as some irons. All of these constituents have characteristics consistent with hard carbonaceous meteoroids. Irons from $H_{\mathrm{class}}=$ C, along with carbonaceous meteors in $H_{\mathrm{class}}=$ D occupy the higher density, fully asteroidal region. The majority of the GEM events were assigned $H_{\mathrm{class}}=$ E and occupy a strongly asteroidal region of Tisserand space, consistent with the low perihelion distance of this shower and its parent body, asteroid (3200) Phaethon \citep{Williams_and_WU_1993}. $H_{\mathrm{class}}$ groupings G, H, and I are concentrated within the cometary domain in both parameters. This is consistent with the cometary meteor showers that dominate these classes, as they exhibit low $T_J$ values and lower activation typical of cometary sources. One example is the LYR meteors from LPC C/1861 G1 Thatcher, which comprised the vast majority of $H_{\mathrm{class}}=$ I. $H_{\mathrm{class}}=$ J, comprised of events from the LYR, GEM, and QUA, may be a subclass of short-period orbits (in the case of LYR type events) or soft material from the two carbonaceous (GEM and QUA) meteor showers. This also suggests that some Lyrids may occupy transitional material strength regimes despite their cometary origin. $H_{\mathrm{class}}=$ K, which consisted of mostly TAH meteors, aligns well with the orbital properties of their JFC parent body, 73/P Schwassmann-Wachmann 3. 

\subsection{Structural Interpretation of $H_{\mathrm{class}}$} 
\label{subsec:macro_classes}

We propose a macro-classification scheme by reviewing the results of $H_{\mathrm{class}}$ applied to known showers (Section~\ref{subsec:showers}) and relative to orbital properties (Section~\ref{subsec:orbElements}). These classes, ranging from the hardest material to softest, are summarized as:

\begin{itemize}
    \item \textbf{$H_{\mathrm{class}}$ A:} Hard carbonaceous material, for example the denser components of the SDAs. These would be Ceplecha's "asteroidal meteors". 
    \item \textbf{$H_{\mathrm{class}}$ C:} High density iron meteoroids.
    \item \textbf{$H_{\mathrm{class}}$ D:} Regular carbonaceous material, similar to Ceplecha class A.
    \item \textbf{$H_{\mathrm{class}}$ E:} Soft carbonaceous material resembling Ceplecha class B or PE class II. An example shower is the Geminids.
    \item \textbf{$H_{\mathrm{class}}$ G:} Regular cometary material comparable to Ceplecha C1 or Ceplecha PE class IIIA.
    \item \textbf{$H_{\mathrm{class}}$ H:} Large regular cometary material, similar to $H_{\mathrm{class}}$ G.
    \item \textbf{$H_{\mathrm{class}}$ I:} Cometary material from LPCs (Ceplecha class C3). An example shower would be the LYR.
    \item \textbf{$H_{\mathrm{class}}$ J:} Very soft cometary material from short-period comets similar to Ceplecha D class. 
    \item \textbf{$H_{\mathrm{class}}$ K:} Very soft and slow cometary material from short-period comets, also similar to Ceplecha D class. The TAH and Draconids (DRA) are examples of showers that fall into this group. 
\end{itemize}

\begin{figure}
    \centering
        \includegraphics[width=\linewidth]{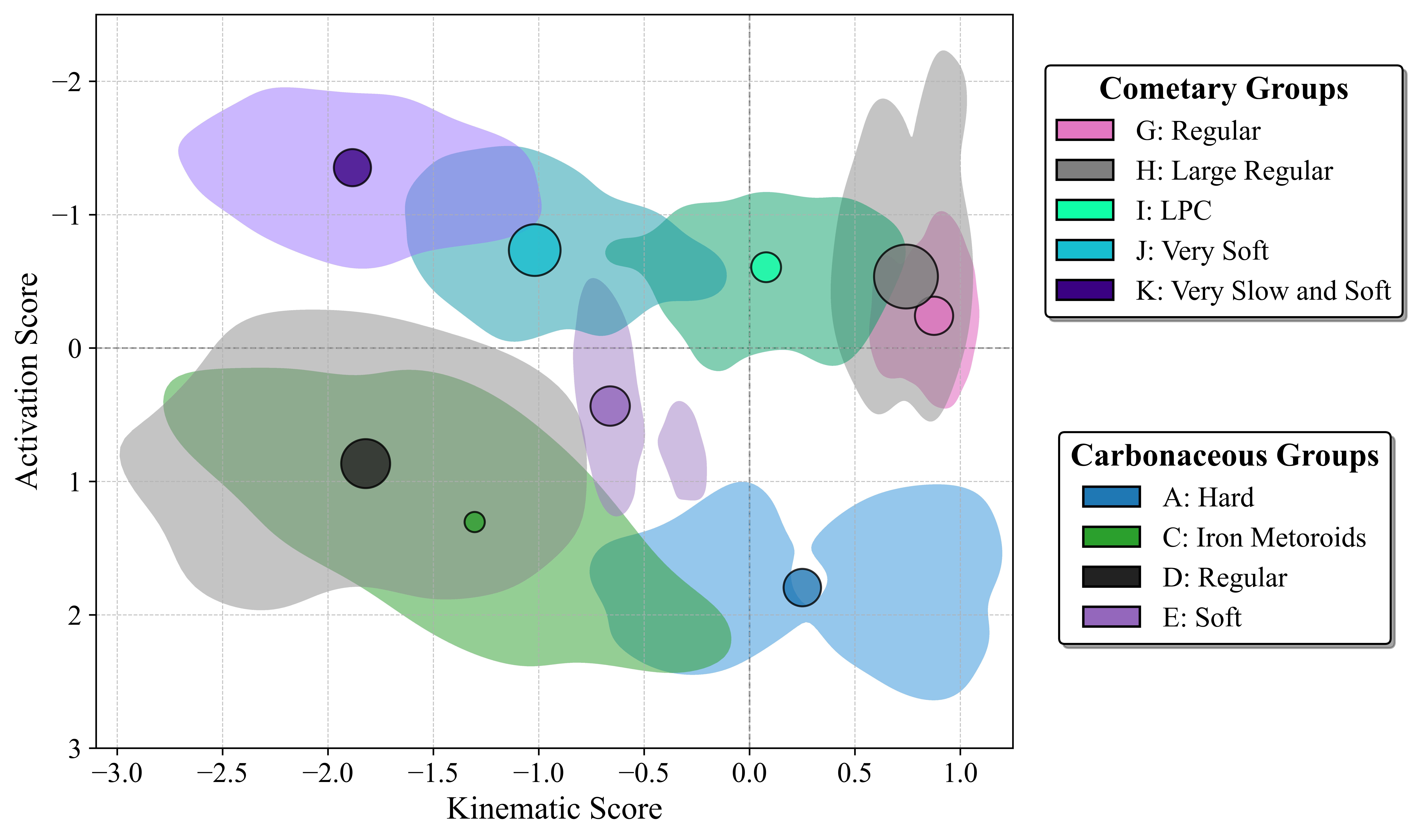}
    \caption{Physical interpretations of the $H_{\mathrm{class}}$ model in factor space. Classes and their associated physical interpretations (Section~\ref{subsec:macro_classes}) are shown in the legend on the right. Median activation score is on the y-axis, where it is inverted to maintain consistency with Figure~\ref{fig:cluster_subfigs}, and median kinematic score is on the x-axis. The size factor is indicated by the size of the circles. For reference, class H has the highest median size normalization score at 1.96 and class C has the lowest median size normalization score at -1.09. Meteors with primary assignment in either class B or F were merged into clusters based on their second highest probability. See Section~\ref{subsec:clustering_results} for details.}
    \label{fig:model_interpretations}
\end{figure}

The Ceplecha classes highlighted above refer to PE classes outlined in \cite{Ceplecha_and_McCrosky_1976} and $K_{b}$ classes from \cite{ceplecha1988}. Figure~\ref{fig:model_interpretations} is a finalized form of the kinematic vs activation figure consistent with the left panel of Figure~\ref{11_cluster_result}, but differing with the removal of clusters B and F. It presents median activation and median kinematic scores for each $H_{\mathrm{class}}$ discussed above along with the physical interpretations in the legend.

\subsection{Comparison of a 9 Cluster Solution to $H_{\mathrm{class}}$}
\label{subsec:9_clusters}

Here, we assess if adopting a 9 cluster model from the outset would have been more effective than the $H_{\mathrm{class}}$ model. To do this, we applied the same analyses described above to a 9 cluster FA–GMM model. This approach allowed us to evaluate whether a 9 cluster solution would provide a comparable representation of the data relative to the $H_{\mathrm{class}}$ model adopted.

\begin{figure}
    \centering
    \includegraphics[width=0.8\linewidth]{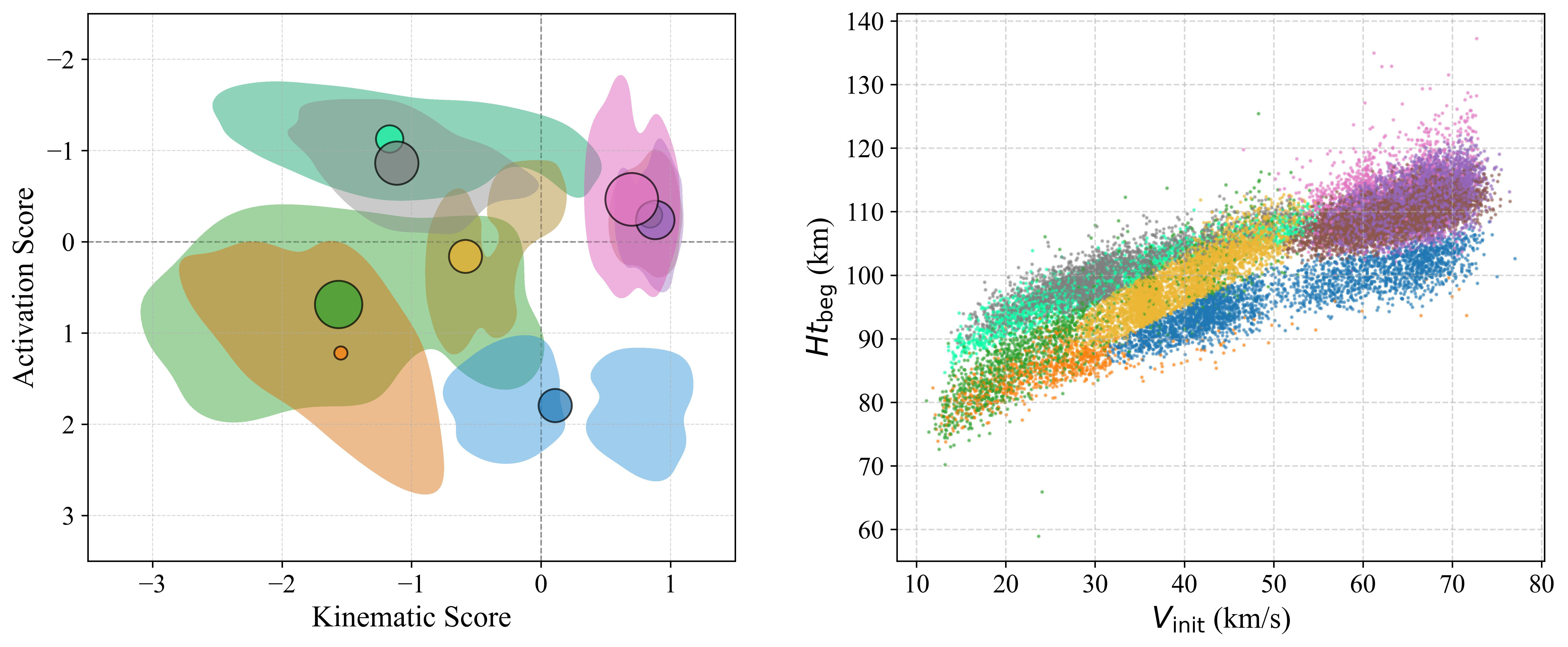}

    \vspace{0.5em}

    \includegraphics[width=0.8\linewidth]{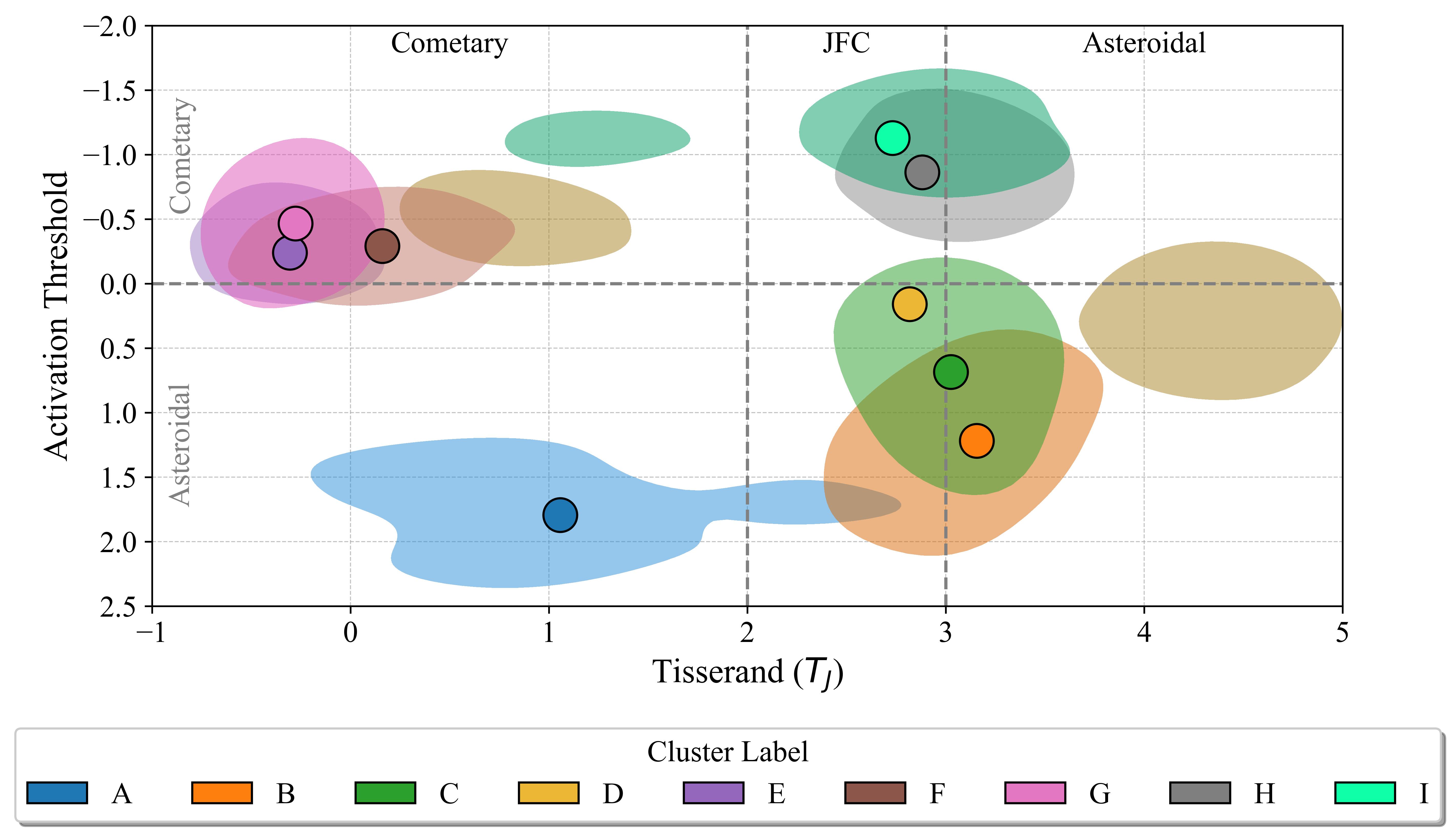}

    \caption{Same as Figure~\ref{fig:cluster_subfigs} (top) and Figure~\ref{fig:tissJplots} (bottom), but for a 9-cluster solution. This solution was not preferred because the three high velocity cometary clusters (E, F, and G in the top right) occupy the same region of orbital space in the bottom panel. A similar structure was revealed in $H_{\mathrm{class}}$, where one cluster was subsequently merged. Clusters D and I also exhibit disconnected contours from their medians, indicating that $H_{\mathrm{class}}$ better resolves physically distinct meteor populations. Colors correspond to cluster assignments in all three panels and the legend is at the bottom.}
    \label{fig:9_cluster}
\end{figure}

The 9 cluster solution produced realistic clustering results consistent with those shown in Figure~\ref{fig:cluster_subfigs} (top right of Figure~\ref{fig:9_cluster}).  Comparing shower clustering between the 9 cluster model and $H_{\mathrm{class}}$ shows that $H_{\mathrm{class}}$ achieves slightly higher overall retention with increases in both the mean (47.6\% to 50.1\%) and median (50.4\% to 52.9\%). Additionally, the range of retention improves, with the minimum increasing from 4.3\% to 13.4\% and the maximum from 82.2\% to 98.7\%, indicating that the $H_{\mathrm{class}}$ solution captures some showers more effectively.

$H_{\mathrm{class}}$ shows retention improvements for several showers, most prominently the LYR, whose retention increases from 4.3\% to 61.5\%. Additional improvements are observed for the TAH, ORI, and GEM. Conversely, retention decreases for some showers under $H_{\mathrm{class}}$, particularly the QUA as well as the SDA and the PER. These results indicate that while the $H_{\mathrm{class}}$ model improves classification for several showers, it also redistributes meteors among clusters in ways that reduce retention for others. A statistical test of the mean change in retention percentage for each shower between the 9 cluster and $H_{\mathrm{class}}$ indicates that the difference is not statistically significant ($t=0.27, p=0.79$).

Evaluation of the 9 cluster model in Tisserand space, as done in Section~\ref{subsec:orbElements}, shows clusters that are less physically interpretable than the $H_{\mathrm{class}}$ groupings. The bottom panel of Figure~\ref{fig:9_cluster} reproduces Figure~\ref{fig:tissJplots} for the 9 cluster model. The three high velocity cometary groupings in the 9 cluster model (E, F, and G) occupy the same region as $H_{\mathrm{class}}$ groups G and H. $H_{\mathrm{class}}$ also identified these same groups and one was dissolved in the final model, highlighting the importance of adding physical interpretations to model results as the number of clusters is increased. Cluster D in the 9 cluster model has two 33\% contours; one that exhibits strong cometary characteristics and another with asteroidal characteristics. Cluster I also shows multiple contours. These split structures suggest a solution with more than 9 clusters better resolves physically distinct populations. Because the shower analysis did not show a statistically significant difference between retention percentage and the 9 cluster solution is less physically interpretable in orbital space, $H_{\mathrm{class}}$ remains the more viable classification framework.

\subsection{Outcomes of Factor Analysis} 
\label{subsec:fa_discussion}

The structure revealed here by FA suggests that meteor compositions can be explained by three factors that serve as umbrellas to describe underlying processes, rather than just a handful of direct observables. Each factor contains at least one term from the $K_{b}$ equation (Equation~\ref{k_b_87}), indicating that $K_{b}$ broadly captures the material structure implied by the physical processes governing meteor behavior. The velocity, atmospheric density, and zenith angle terms that are used in $K_{b}$ also comprise several of our most significant features.  However, our results indicate that there is a depth to this picture that a three parameter model may be missing. By incorporating a broader set of observables, FA uncovered a latent structure that organizes meteoroid classification across the 13 parameters considered. This structure highlights the role kinematic, activation threshold, and size-related processes play in governing meteor ablation.

A key reason FA performs better than PCA here is that many of the measured features are inherently correlated (Figure~\ref{corr_matrix}). For example, beginning atmospheric density is always tied to the beginning height itself, and the two height and two velocity terms are intrinsically linked and will always co-vary for a given event. PCA treats these correlations as shared variance to be projected along orthogonal axes, while FA explicitly models them as the result of shared latent physical drivers. In other words, FA assumes that features co-vary because they are linked by common physical processes, such as entry speed, ablation efficiency, or mass/size scaling, rather than requiring the axes to be orthogonal.

FA structure implies that meteoroid behavior is governed by coupled processes that simultaneously affect observed parameters. A lower dimensional model like $K_{b}$ produces the broad separation between cometary and carbonaceous material, but it does not fully capture how combinations of features interact. FA, by contrast, provides a richer map of how these dependencies manifest in the data which opens the possibility of identifying subtle subpopulations and physical regimes that would otherwise remain hidden in traditional classifications.

\subsection{Implications of Bias by Using a Semi-Qualitative Approach} \label{subsec:bias}

Our workflow intentionally utilized a semi-qualitative approach. We used $S$ to provide a quantitative assessment of cluster quality, but clusters were treated as physically realistic only if they aligned with established expectations from $K_{b}$, as seen in Figure~\ref{kb_ht_vel}. While this approach ensured that our classifications maintained well established patterns in the data, it also introduced an inherent bias. Clusters were only deemed realistic when they aligned with current models rather than allowed to stand independently.

This step was partly motivated by a fundamental limitation of the GMM. It assumes that cluster structure is approximately Gaussian when meteor populations are likely non-Gaussian. Because the true distribution of clusters is not known prior to fitting, statistical groupings must be interpreted in a physical context to determine whether they correspond to meaningful meteoroid populations. The activation threshold itself is derived directly from the observed parameters and is therefore not expected to depend on biases introduced by the clustering algorithm.

We aimed to ensure that patterns identified by the clustering process represented reality, especially given that our clustering operated in a 13-dimensional feature space. This added complexity expands on the 3-dimensional space utilized by $K_{b}$. Thus, we present $H_{\mathrm{class}}$ as a data-driven extension of $K_{b}$. A more complex model allows for new interpretations, but requiring it to resemble current understandings also risks reinforcing prior assumptions and overlooking subtle structure in the data. By judging clusters on how well they align with existing expectations, we may be limiting the discovery of unexpected structures in the data. Interpreting a 13-dimensional feature space through only two dimensions, $Ht_\mathrm{beg}$ and $V_\mathrm{init}$, may have limited what can be learned.

Clear cluster separations were found in $Ht_\mathrm{beg}$ and $V_\mathrm{init}$ space. Given that a higher dimensional feature space is encoded here, such separability might imply that predominant meteoroid properties can be inferred from only these two parameters. The class that a meteor belongs to reflects defining characteristics, and the clear separability we see among these two dimensions suggests that $Ht_\mathrm{beg}$ and $V_\mathrm{init}$ imply meaningful physical behavior. Confirming the extent to which just these two variables can inform meteor behavior would require additional inquiry.

A related concern lies with whether the distinction between high-velocity and low-velocity cometary meteors (and the subclusters therein) reflect a true difference in meteoroid properties or is simply an artifact of the clustering process. Models consistently separated these two groups across our tests (Figure~\ref{fig:rej_methods}), suggesting this may be a meaningful division not captured in a model such as $K_{b}$. Further analysis would be needed to determine whether this reflects real physical structure or methodological bias.

Without the qualitative constraints we imposed, the resulting clusters might have revealed alternative patterns. A purely data-driven classification scheme could highlight hidden relationships between features or divide correlated features into unexpected subgroups. Such patterns could challenge assumptions about material strength or the behavior of showers from the same parent body. The results shown in Figures \ref{fig:rej_methods} and \ref{fig:hdbscan} were not used further because their clusters were deemed unrealistic. However, the patterns revealed by these methods may point to structure in the data that deserve follow-up analysis. Section~\ref{sec:rej_methods} outlines the techniques we tried but ultimately discarded. We present this framework as a guide for others seeking to explore these and possibly different model combinations to uncover hidden patterns in meteor data.

\subsection{Code Availability} \label{code}

\ifanonym   
    Code to implement this model will be made available, this section omitted for double-blind review.
\else
    \texttt{Python} code developed as a part of this project are available on GitHub at \url{https://github.com/sammmelg/hcmm}. The available scripts convert GMN data into .csv files, which can then be loaded by the modeling script. The modeling script applies the analytical model described in Appendix \ref{model_form} with the coefficients from Appendix \ref{model_params} for the 11 cluster model and returns the resulting $H_{\mathrm{class}}$ for each event. This code is made available under an MIT open source license and is intended primarily for applying the current model to future datasets. Users are encouraged to modify and enhance these functions.
\fi

\section{Conclusion} \label{sec:conclusion}

We developed a classification framework that uses ML to identify compositional structure in meteoroid observations from the LO-CAMS network in Arizona and the wider GMN. Our goal was to determine whether physically meaningful patterns could be extracted directly from observed features while expanding on traditional classification schemes. After testing a range of normalization methods, dimensionality-reduction techniques, and clustering algorithms, we identified that using Factor Analysis and a Gaussian Mixture Model produced the most stable and physically interpretable results.

The 13-dimensional feature space captured a broader set of physical behavior than existing approaches such as the $K_{b}$ parameter. The resulting cluster structures reproduced known groupings while revealing additional substructure. Applying our final model to nine well-studied meteor showers aided the physical interpretation of the clusters, showed that they are consistent with expected material properties and dynamical origins, and also highlighted differences among showers from the same parent body. These results demonstrate that ML methods are a worthwhile tool for understanding meteoroid strength and classifying events accordingly. This work establishes a framework that can be applied to future meteor observations. Future efforts can incorporate measurement uncertainties and explore alternative or expanded feature sets.

Our main findings are as follows:

\begin{itemize}
    \item We built a ML workflow using 13 directly observed parameters to classify meteoroids based on inferred composition and expand upon the $K_{b}$ parameter.
    \item We tested multiple combinations of normalization schemes, dimensionality-reduction methods (FA, PCA, UMAP), and clustering algorithms (GMM, BGMM, KMeans, Agglomerative, Birch, Spectral, and HDBSCAN).
    \item Factor Analysis combined with a Gaussian Mixture Model produced the most physically interpretable results with a methodology that is readily extensible to other data sets.
    \item The latent factors recovered by FA correspond to meaningful physical drivers: meteoroid kinematics, activation thresholds, and size/viewing geometry related behavior.
    \item The 3, 6, and 11 cluster models revealed progressively finer structure, ranging from broad physical regimes to more detailed subdivisions tied to meteor behavior in Earth's atmosphere.
    \item We defined $H_{\mathrm{class}}$ which provides a physically motivated hardness nomenclature that categorizes meteoroids by their resistance to ablation and fragmentation using activation threshold scores.
    \item Applying $H_{\mathrm{class}}$ to nine well-studied meteor showers facilitated the interpretation of the material properties associated with the resulting clusters.
    \item Mapping clusters into orbital element space demonstrated consistency between the derived physical classes and expected cometary, JFC, and asteroidal populations.
    \item $H_{\mathrm{class}}$ groupings outlined a structural spectrum of meteoroid materials, from the densest carbonaceous and iron-rich meteoroids to the softest cometary populations.
    \item Our semi-qualitative evaluation may have introduced bias toward known physical expectations, but also ensured that the final model aligns with established behavior.
    \item The discarded methods (Section~\ref{sec:rej_methods}) still revealed structure that may warrant follow-up, especially for identifying subtle or unexpected relationships.
\end{itemize} 

%% Please use the acknowledgment and contribution environments. This will 
%% be anonomyized when the "anonymous" style option is used. 
\section*{Acknowledgments}
\ifanonym
    Acknowledgments omitted for double-blind review.

\else
    We acknowledge the two anonymous reviewers who examined our work. Their comments made our work more robust and we appreciate their thoughtful reviews. 
    
    SH acknowledges E. Peña-Asensio for a helpful discussion regarding this work and the sharing of his HDBSCAN implementation at the joint DPS/EPSC Conference in Helsinki, Finland. SH also acknowledges NAU Assistant Professor Dr. Lan Zhang and classmate Comfort Norteye for inspiring this project in a ML class at NAU in Fall '24 along with their assistance in the initial iterations of this work.

    We are deeply grateful to the students, collaborators, and station hosts who have helped support the LO-CAMS network including: Peter Jenniskens, Megan Gialluca, Solvay Blomquist, Robert Schottland, Matt Francis, Matt Cheselka, Bob Broffel, staff at the Navajo Nation Museum, and John Glitsos at the White Mountain Nature Center.

    SH and NM acknowledge support awarded by the National Science Foundation to CAREER grant No. AST1944827 (PI Moskovitz).

    DV was partially funded by the NASA Meteoroid Environment Office under cooperative agreement 80NSSC24M0060.
\fi

\section*{Software}
\texttt{matplotlib} \citep{Hunter:2007}, \texttt{numpy} \citep{harris2020array}, \texttt{pandas} \citep{mckinney-proc-scipy-2010, reback2020pandas}, {\texttt{scikit-learn} \citep{scikit-learn}, \texttt{seaborn} \citep{Waskom2021}, \texttt{wmpl} \citep{Vida_2019}}

\section*{Declaration of generative AI and AI-assisted technologies in the manuscript preparation process.}

During the preparation of this work the author(s) used ChatGPT 5.2 to help refine sentence structure, paragraph flow, and avoid word repetition. After using this tool/service, the author(s) reviewed and edited the content as needed and take(s) full responsibility for the content of the published article.
\printcredits

\bibliographystyle{cas-model2-names}

\bibliography{ms_class_mlBib}

\appendix
\clearpage

\section*{Appendix}
\section{Cluster Medians in Factor Space for the 3, 6, and 11 Cluster Models}
\label{cluster_medians}

This section summarizes median values per cluster for each of the Kinematic, Activation Threshold, and Size Normalization factors. Values are presented to 8 decimal places to aid those intending to replicate this work.

\begin{center}
\small
\captionof{table}{Cluster medians in factor space for the 3 cluster FA-GMM model.}
\label{tab:3_cluster_medians}
\begin{tabular}{lccc}
\toprule
Cluster & Kinematic & Activation Threshold & Size \\
\midrule
A & -0.42956258 & 1.39340746 & -0.28589612 \\
B & 0.86228426 & -0.28393377 & -0.22857780 \\
C & -0.70302290 & -0.59917033 & 0.25401484 \\
\bottomrule
\end{tabular}
\end{center}

\begin{center}
\small
\captionof{table}{Cluster medians in factor space for the 6 cluster FA-GMM model.}
\label{tab:6_cluster_medians}

\begin{tabular}{lccc}
\toprule
Cluster & Kinematic & Activation Threshold & Size \\
\midrule
A & -0.02724807 & 1.80843043 & -0.21769527 \\
B & -1.72731865 & 0.52123912 & 0.82542614 \\
C & -0.51879574 & 0.15118178 & -0.26005742 \\
D & 0.86532942 & -0.27449808 & 0.54792951 \\
E & 0.85368772 & -0.27626497 & -0.67536664 \\
F & -1.13836976 & -0.94203625 & 0.31189496 \\
\bottomrule
\end{tabular}
\end{center}

\begin{center}
\small
\captionof{table}{Cluster medians in factor space for the 11 cluster FA-GMM model.}
\label{tab:11_cluster_medians}

\begin{tabular}{lccc}
\toprule
Cluster & Kinematic & Activation Threshold & Size \\
\midrule
A  & 0.13677511  & 1.80460939  & -0.27769966 \\
B  & -0.08350765 & 1.33229519  & 1.79413352  \\
C  & -1.41002751 & 1.17883405  & -1.26555344 \\
D  & -1.90114777 & 0.85928157  & 0.46474272  \\
E  & -0.66392613 & 0.41820220  & -0.17215429 \\
F  & 0.88033229  & -0.25150153 & -0.73644813 \\
G  & 0.84995456  & -0.26267695 & 0.37233834  \\
H  & 0.74247383  & -0.57709124 & 1.96876859  \\
I  & -0.01300870 & -0.59930166 & -0.51347819 \\
J  & -1.04069637 & -0.74853774 & 0.75524533  \\
K  & -1.88587367 & -1.35429963 & -0.28348929 \\
\bottomrule
\end{tabular}
\end{center}

\clearpage
\section{Shower Cluster Membership and Retention Statistics}\label{sankey_stats}

\begin{center}
\small
\captionof{table}{Shower retention statistics and cluster membership counts for the $H_{\mathrm{class}}$ model. The $Pre$ column indicates the total number of shower meteors present before applying confident assignment cutoffs, $Post$ is the number of shower meteors confidently assigned to a cluster. Retention percentage represents the proportion of shower events confidently assigned to a cluster. Our confident assignment threshold was those events with a $\geq 80\%$ probability of cluster assignment.}
\label{tab:cluster_counts_11}
\begin{tabular}{lccccccccccccc}
\toprule
IAU & Pre & Post & Ret. (\%) & A & C & D & E & G & H & I & J & K \\
\midrule
TAH & 1115 & 1101 & 98.7 & 0 & 0 & 2 & 0 & 0 & 0 & 0 & 0 & 1099 \\
LYR & 6275 & 3860 & 61.5 & 22 & 2 & 2 & 0 & 0 & 95 & 3575 & 164 & 0 \\
ETA & 19480 & 11211 & 57.6 & 48 & 0 & 0 & 0 & 9286 & 1790 & 87 & 0 & 0 \\
ORI & 30841 & 17317 & 56.1 & 68 & 0 & 0 & 0 & 15894 & 850 & 505 & 0 & 0 \\
GEM & 47449 & 25079 & 52.9 & 33 & 4 & 735 & 24142 & 0 & 0 & 1 & 164 & 0 \\
Iron & 19991 & 10187 & 51.0 & 1522 & 7175 & 723 & 405 & 0 & 0 & 13 & 0 & 349 \\
PER & 72094 & 27794 & 38.6 & 71 & 3 & 0 & 0 & 20625 & 5185 & 1910 & 0 & 0 \\
SDA & 16586 & 3446 & 20.8 & 1193 & 35 & 228 & 1866 & 0 & 0 & 73 & 51 & 0 \\
QUA & 8841 & 1183 & 13.4 & 109 & 0 & 142 & 458 & 0 & 0 & 284 & 190 & 0 \\
\midrule
Total & 222672 & 101178 & 45.4 & 3066 & 7219 & 1832 & 26871 & 45805 & 7920 & 6448 & 569 & 1448 \\
\bottomrule
\end{tabular}
\end{center}

\begin{center}
\small
\captionof{table}{Cluster membership percentages for each shower under the $H_{\mathrm{class}}$ model.}
\label{tab:cluster_percentages_11}

\begin{tabular}{lccccccccc}
\toprule
IAU & A & C & D & E & G & H & I & J & K \\
\midrule
TAH  & 0.0 & 0.0 & 0.2 & 0.0 & 0.0 & 0.0 & 0.0 & 0.0 & 99.8 \\
LYR  & 0.6 & 0.1 & 0.1 & 0.0 & 0.0 & 2.5 & 92.6 & 4.2 & 0.0 \\
ETA  & 0.4 & 0.0 & 0.0 & 0.0 & 82.8 & 16.0 & 0.8 & 0.0 & 0.0 \\
ORI  & 0.4 & 0.0 & 0.0 & 0.0 & 91.8 & 4.9 & 2.9 & 0.0 & 0.0 \\
GEM  & 0.1 & 0.0 & 2.9 & 96.3 & 0.0 & 0.0 & 0.0 & 0.7 & 0.0 \\
Iron & 14.9 & 70.4 & 7.1 & 4.0 & 0.0 & 0.0 & 0.1 & 0.0 & 3.4 \\
PER  & 0.3 & 0.0 & 0.0 & 0.0 & 74.2 & 18.7 & 6.9 & 0.0 & 0.0 \\
SDA  & 34.6 & 1.0 & 6.6 & 54.1 & 0.0 & 0.0 & 2.1 & 1.5 & 0.0 \\
QUA  & 9.2 & 0.0 & 12.0 & 38.7 & 0.0 & 0.0 & 24.0 & 16.1 & 0.0 \\
\bottomrule
\end{tabular}
\end{center}

\section{Mathematical Formalism of An Analytical Model} \label{model_form}

A primary objective of this work was to provide a model that can be readily applied to future meteor observations. In this section, we provide the mathematical formalism that underlies the model. Appendix \ref{model_params} provides coefficients for all terms derived from the training dataset that can be implemented using these equations to reproduce the model.

\subsection{Min/Max Normalization} \label{subsubsec:norm}

Min/Max scaling is applied to each event so that all features lie within the range [a,b] = [-1, 1]. For a single feature value $x_{m,n}$ (event $m$, feature $n$), the normalized quantity is:

\begin{equation}
    x'_{m, n} = \frac{x_{m,n} - x_{n, \min}}{x_{n, \max} - x_{n, \min}}(b - a) + a
    \label{equ:normalize1}
\end{equation}

Where $x_{n, \min}$ and $x_{n, \max}$ represent the minimum and maximum values of feature $n$ and $a=-1$ and $b=1$ to specify the target range. We define $D$ as an $(n \times n)$ diagonal scaling matrix for each feature, $m$ as an $(n \times 1)$ matrix of minimum values, and $\beta$ as an $(n \times 1)$ shift vector that moves each feature into range. We now have:

\begin{equation}
    D = \mathrm{diag}\!\left(\frac{2}{x_{n,\max} - x_{n,\min}}\right); \qquad 
    m = \begin{bmatrix}
        x_{1, \min} \\ 
        x_{2, \min} \\ 
        \vdots \\ 
        x_{n, \min}
        \end{bmatrix}; \qquad
    \beta = -1 - Dm
    \label{equ:normalize2}
\end{equation}

The Min/Max normalization for any event can be written as:

\begin{equation}
    x'_{m} = Dx + \beta
    \label{equ:normalize3}
\end{equation}

Where $x'$ represents the normalized $(n \times 1)$ feature vector for any event. Table~\ref{tab:minmax_features} in the Appendix shows minimum and maximum values, $x_{n, \min}$ and $x_{n, \max}$, for each feature from the LO-CAMS dataset.

\subsection{Factor Analysis} \label{subsec:fa_model}

This section provides the mathematical details underlying the computation of the factor score matrix 
$F$ referenced in Section~\ref{subsec:fa}, including the assumptions and linear operations used to estimate $F$ from the normalized feature matrix $X'$. Under our model, FA assumes that each meteor event is modeled as a linear combination of $j=3$ latent factors as expressed in Equation~\ref{Matrix}. A standard assumption in FA is that the latent factor score vector for each event follows a multivariate normal distribution with zero mean and identity covariance:

\begin{equation}
    f_{m} \sim \mathcal{N}(0, I_{3})
    \label{equ:fa1}
\end{equation}

The factor loadings $W$ and feature-specific noise variances $\Psi$ are estimated from the training data by maximizing the FA likelihood, after which the Varimax rotation is applied to the loadings. This rotation is an input parameter to the \texttt{FactorAnalysis} function and is incorporated directly into the factor loadings $W$, so we will not go through the mathematical application of the transformation. We suggest \cite{Kaiser1958TheVC} as a reference for further reading on this topic. The bias term $M$ in Equation~\ref{Matrix} corresponds to the feature-wise mean vector $\mu$ of the normalized data matrix $X'$, replicated across all events, such that $M=1_{m}\mu^{\top}$. To compute factor scores from the observed, normalized features, FA first centers the data by subtracting $\mu$, weights each feature by its inverse noise variance, and then projects these weighted observations onto the factor loadings. Mathematically this projection is written as:

\begin{equation}
    W^{\top}\Psi^{-1}\left(X'-\mu\right)
    \label{equ:fa2}
\end{equation}

Where $\Psi = \mathrm{diag}(\psi_1, \psi_2, \ldots, \psi_n)$ is the diagonal noise covariance matrix. Its inverse, $\Psi^{-1}$, acts as a precision matrix, downweighting features with large noise variance and upweighting features with small noise variance. FA first forms this precision-weighted projection of the centered observations into factor space, providing a measure for how strongly each observation supports the latent factors $f_m$. 

Under the prior assumption from Equation~\ref{equ:fa1}, factor scores are not taken directly from this projection. Instead, the prior regularizes the projection to prevent factor magnitudes from becoming arbitrarily large. This regularization enters through the matrix:

\begin{equation}
    \left(W^{\top}\Psi^{-1}W+I\right)^{-1}
    \label{equ:fa3}
\end{equation}

Which balances the information provided by the data with the unit-variance prior on the latent factors and yields a stable estimate. To obtain the final factor scores, FA combines this regularization with the precision-weighted projection in Equation~\ref{equ:fa2}, yielding the linear estimator $\Omega$:

\begin{equation}
    \Omega = \left(W^{\top} \Psi^{-1} W + I_3 \right)^{-1} W^{\top} \Psi^{-1}.
    \label{equ:fa4}
\end{equation}

Applying this linear operator to the centered observations gives the matrix of factor scores:

\begin{equation}
    F = \Omega (X' - \mu).
    \label{equ:fa5}
\end{equation}

\subsection{Gaussian Mixture Model} \label{subsec:gmm_model}

Here we describe how posterior cluster probabilities and hard class assignments are computed for individual events using the fitted GMM parameters. For any given factor vector, $f_{m}$, the probabilitiy density function (PDF) for cluster $k$ under a 3 factor model is defined as:

\begin{equation}
    \mathcal{N}(f_{m}|\mu_{k}, \Sigma_{k}) = \frac{1}{\left(2\pi\right)^{\frac{3}{2}}|\Sigma_{k}|^{\frac{1}{2}}}exp^{\left[-\frac{1}{2}\left(f_{m}-\mu_{k}\right)^{\top}\Sigma_{k}^{-1}\left(f_{m}-\mu_{k}\right)\right]}
    \label{equ:gmm1}
\end{equation}

Where $\mu_{k}$ is the mean vector for each cluster from the fitted GMM on LO-CAMS data and $\Sigma_{k}$ is the covariance matrix for each cluster. The weighted PDF, which quantifies the contribution of cluster $k$ to the event, for each cluster from the fitted model:

\begin{equation}
    p(f_{m}| z= k) = \pi_{k}\mathcal{N}(f_{m}|\mu_{k}, \Sigma_{k})
    \label{equ:gmm2}
\end{equation}

Where $\pi_{k}$ represents the mixture weights for each cluster. We can then compute the total weighted PDF, which is how well all clusters combined explain any particular event:

\begin{equation}
    p(f_{m}) = \sum_{l=1}^{K}\pi_{l}\mathcal{N}(f_{m}|\mu_{l}, \Sigma_{l})
    \label{equ:gmm3}
\end{equation}

The posterior probability that any event, $m$, belongs to cluster $k$ after observing its factor vector, $f_{m}$, can now be computed:

\begin{equation}
    \gamma_{mk} = p(z_{m} = k | f_{m}) = \frac{p(f_{m}| z= k)}{p(f_{m})} 
    \label{equ:gmm4}
\end{equation}

Finally, the $H_{\mathrm{class}}$ can be found by determining the cluster that has the highest probability of assignment:

\begin{equation}
    H_{\mathrm{class}} = \arg\max_{k} \gamma_{mk}
    \label{equ:gmm5}
\end{equation}

Substituting Equations \ref{equ:gmm2} and \ref{equ:gmm3}, into Equation~\ref{equ:gmm4} yields one expression that more explicitly describes the entire model:

\begin{equation}
     H_{\mathrm{class}} = \arg\max_{k} \frac{\pi_{k}\mathcal{N}(f_{m}|\mu_{k}, \Sigma_{k})}{\sum_{l=1}^{K}\pi_{l}\mathcal{N}(f_{m}|\mu_{l}, \Sigma_{l})}
    \label{equ:gmm6}
\end{equation}

\clearpage
\section{Model Parameters} \label{model_params}

We provide necessary coefficients to implement the 11 cluster model. Coefficients are provided to a precision of 8 decimal places to aid in reproducing the derived models.

\subsection{Normalization Parameters} \label{model_params_1}

\begin{center}
\begin{minipage}{0.85\linewidth}
\centering
\captionof{table}{Min/Max Scaling Values}
\label{tab:minmax_features}

\begin{tabular}{lrr}
\hline
Scaled Feature                & Min Value   & Max Value   \\
\hline
$\log_{10}(E_\mathrm{beg})$          & -0.80400014 &  2.84297654 \\
$\log_{10}(\rho_\mathrm{beg})$       & -8.37307706 & -3.48330594 \\
$\log_{10}(Ht_\mathrm{beg})$         &  1.77013815 &  2.13747786 \\
$\log_{10}(Ht_\mathrm{end})$         &  1.67034514 &  2.05708310 \\
$\log_{10}(V_\mathrm{init})$         &  1.04601261 &  1.88666565 \\
$\log_{10}(V_\mathrm{avg})$          &  1.01178521 &  1.86380502 \\
$\log_{10}(L_\mathrm{trail})$        &  0.38372814 &  2.31833898 \\
$t_\mathrm{dur}$                     &  0.20000000 &  4.80000000 \\
$\cos(\theta_z)$                     &  0.04582851 &  0.99998911 \\
$M_\mathrm{abs}$                     & -9.39000000 &  2.83000000 \\
$F$                                  &  0.00000000 &  1.00000000 \\
$\log_{10}(m)$                       & -5.80410035 & -0.79588002 \\
$\log_{10}(a_\mathrm{decel})$        & -4.61278386 &  1.67488333 \\
\hline
\end{tabular}

\vspace{0.5em}
{\footnotesize Minimum and maximum values used for Min/Max normalization of all 13 features.}
\end{minipage}
\end{center}

\subsection{Factor Analysis Parameters} \label{model_params_2}

\begin{center}
\begin{minipage}{0.95\linewidth}
\centering
\captionof{table}{Factor Analysis Parameters: Feature Means, Factor Loadings, and Noise Terms.}
\label{tab:fa_params}

\begin{tabular}{lccccc}
\hline
Feature & Mean & Kinematics & Activation Threshold & Size Normalization & $\psi_n$ \\
\hline
$E_\mathrm{beg}$        & $-0.28092403$ & $0.08455906$ & $0.16540866$ & $0.01064679$ & $5.66713919\times10^{-3}$ \\
$\rho_\mathrm{beg}$     & $-0.25107786$ & $0.23780462$ & $0.13742304$ & $-0.03751511$ & $6.31336263\times10^{-4}$ \\
$Ht_\mathrm{beg}$       & $0.32499514$  & $-0.17516092$ & $-0.10347116$ & $0.02902104$ & $1.01847910\times10^{-5}$ \\
$Ht_\mathrm{end}$       & $0.47153212$  & $-0.18134596$ & $-0.06159221$ & $-0.03750781$ & $1.00689688\times10^{-2}$ \\
$V_\mathrm{init}$       & $0.48262660$  & $-0.40466890$ & $-0.00494070$ & $0.00646431$ & $1.16785196\times10^{-5}$ \\
$V_\mathrm{avg}$        & $0.51772416$  & $-0.40018603$ & $-0.00860136$ & $0.01195469$ & $2.90494937\times10^{-4}$ \\
$L_\mathrm{trail}$      & $-0.07208113$ & $-0.05406439$ & $-0.06148199$ & $0.19556488$ & $4.67357712\times10^{-5}$ \\
$t_\mathrm{dur}$        & $-0.87424480$ & $0.06972400$  & $-0.03218973$ & $0.10761494$ & $2.69195025\times10^{-3}$ \\
$\theta_z$              & $0.36454656$  & $0.06875790$  & $-0.02102660$ & $-0.24232600$ & $1.30714148\times10^{-1}$ \\
$M_\mathrm{abs}$        & $0.48804189$  & $0.08187245$  & $-0.00567357$ & $-0.06364635$ & $1.95302567\times10^{-2}$ \\
$F$                     & $0.17179600$  & $-0.00980684$ & $-0.06964115$ & $0.05369203$ & $1.10955420\times10^{-1}$ \\
$m$                     & $-0.37565354$ & $0.09872270$  & $-0.01101272$ & $0.12692187$ & $1.82986733\times10^{-2}$ \\
$a_\mathrm{decel}$      & $0.56331339$  & $-0.08597392$ & $0.03648469$  & $-0.09204429$ & $1.42251267\times10^{-2}$ \\
\hline
\end{tabular}

\vspace{0.5em}
{\footnotesize Means, factor loadings, and feature noise variances ($\psi_n$) obtained from the 3 factor FA model after Min/Max normalization to the range $[-1,1]$.}
\end{minipage}
\end{center}

\subsection{Gaussian Mixture Model Parameters} \label{model_params_3}
\begin{center}
\begin{minipage}{0.98\linewidth} % slightly wider, this table is big
\centering
\captionof{table}{11 Cluster Gaussian Mixture Model Weights, Means, and Covariances.}
\label{tab:gmm_params11}

\begin{tabular}{crrrc}
\hline
Class & $\pi_k$ & $\mu_{k,1}$ & $\mu_{k,2}$ & $\mu_{k,3}$ \\
      &         & \multicolumn{3}{c}{$\Sigma_k$} \\
\hline
A  & 0.08434407 & $-0.19829437$ & $1.80178582$ & $-0.29261296$ \\
 & & \multicolumn{3}{c}{
$\begin{bmatrix}
 4.09163595\times10^{-1} & -2.09358089\times10^{-2} & -1.93629015\times10^{-1} \\
 -2.09358089\times10^{-2} & 2.78398159\times10^{-1} & 4.95791222\times10^{-2} \\
 -1.93629015\times10^{-1} & 4.95791222\times10^{-2} & 3.69724945\times10^{-1}
\end{bmatrix}$ } \\
\hline
B  & 0.03632081 & $0.04254277$ & $0.97892327$ & $1.41754598$ \\
 & & \multicolumn{3}{c}{
$\begin{bmatrix}
 4.54888467\times10^{-1} & -2.68215209\times10^{-1} & 1.09873131\times10^{-1} \\
 -2.68215209\times10^{-1} & 1.70876411 & -3.71064363\times10^{-1} \\
 1.09873131\times10^{-1} & -3.71064363\times10^{-1} & 1.10899452
\end{bmatrix}$ } \\
\hline
C  & 0.03772070 & $1.26942937$ & $1.22519887$ & $-0.94574602$ \\
 & & \multicolumn{3}{c}{
$\begin{bmatrix}
 7.88655009\times10^{-1} & -3.88182591\times10^{-1} & -7.36928687\times10^{-1} \\
 -3.88182591\times10^{-1} & 5.44790577\times10^{-1} & 5.20428194\times10^{-1} \\
 -7.36928687\times10^{-1} & 5.20428194\times10^{-1} & 1.22536367
\end{bmatrix}$ } \\
\hline
D & 0.05001716 & $1.76396772$ & $0.88472734$ & $0.40652344$ \\
 & & \multicolumn{3}{c}{
$\begin{bmatrix}
 5.70409728\times10^{-1} & 2.78184940\times10^{-3} & 2.77196451\times10^{-2} \\
 2.78184940\times10^{-3} & 6.43335648\times10^{-1} & -4.40125717\times10^{-2} \\
 2.77196451\times10^{-2} & -4.40125717\times10^{-2} & 1.06573178
\end{bmatrix}$ } \\
\hline
E & 0.09534943 & $0.61781199$ & $0.44382246$ & $-0.18909387$ \\
 & & \multicolumn{3}{c}{
$\begin{bmatrix}
 4.01795791\times10^{-2} & -5.80259811\times10^{-2} & -2.74640646\times10^{-2} \\
 -5.80259811\times10^{-2} & 2.86310476\times10^{-1} & 5.29328318\times10^{-2} \\
 -2.74640646\times10^{-2} & 5.29328318\times10^{-2} & 6.30005184\times10^{-1}
\end{bmatrix}$ } \\
\hline
F  & 0.19519483 & $-0.85136621$ & $-0.24148004$ & $-0.70421087$ \\
 & & \multicolumn{3}{c}{
$\begin{bmatrix}
 2.46707022\times10^{-2} & -1.23853834\times10^{-3} & -1.96850265\times10^{-2} \\
 -1.23853834\times10^{-3} & 2.47036363\times10^{-1} & 7.39676488\times10^{-2} \\
 -1.96850265\times10^{-2} & 7.39676488\times10^{-2} & 1.90197948\times10^{-1}
\end{bmatrix}$ } \\
\hline
G  & 0.19300340 & $-0.82664826$ & $-0.24246533$ & $0.30471965$ \\
 & & \multicolumn{3}{c}{
$\begin{bmatrix}
 2.77301876\times10^{-2} & 6.68284232\times10^{-3} & -1.22476665\times10^{-2} \\
 6.68284232\times10^{-3} & 2.58858876\times10^{-1} & 2.36368558\times10^{-2} \\
 -1.22476665\times10^{-2} & 2.36368558\times10^{-2} & 4.27155224\times10^{-1}
\end{bmatrix}$ } \\
\hline
H  & 0.05098616 & $-0.76606149$ & $-0.54692038$ & $1.52777391$ \\
 & & \multicolumn{3}{c}{
$\begin{bmatrix}
 3.61636739\times10^{-2} & 3.21434143\times10^{-3} & 1.04219699\times10^{-3} \\
 3.21434143\times10^{-3} & 6.97404530\times10^{-1} & 2.42853395\times10^{-1} \\
 1.04219699\times10^{-3} & 2.42853395\times10^{-1} & 1.19989329
\end{bmatrix}$ } \\
\hline
I  & 0.10158541 & $0.00486575$ & $-0.56905512$ & $-0.51093435$ \\
 & & \multicolumn{3}{c}{
$\begin{bmatrix}
 2.00612222\times10^{-1} & -2.27309910\times10^{-2} & 2.24217367\times10^{-2} \\
 -2.27309910\times10^{-2} & 2.26113740\times10^{-1} & 8.06802450\times10^{-2} \\
 2.24217367\times10^{-2} & 8.06802450\times10^{-2} & 5.31648330\times10^{-1}
\end{bmatrix}$ } \\
\hline
J  & 0.09585493 & $0.90835717$ & $-0.71277956$ & $0.63000584$ \\
 & & \multicolumn{3}{c}{
$\begin{bmatrix}
 3.44744611\times10^{-1} & -1.28457018\times10^{-1} & -2.68145784\times10^{-2} \\
 -1.28457018\times10^{-1} & 2.96660472\times10^{-1} & 2.93133697\times10^{-2} \\
 -2.68145784\times10^{-2} & 2.93133697\times10^{-2} & 5.88113076\times10^{-1}
\end{bmatrix}$ } \\
\hline
K  & 0.05962311 & $1.63328065$ & $-1.21364997$ & $-0.01962040$ \\
 & & \multicolumn{3}{c}{
$\begin{bmatrix}
 3.97853037\times10^{-1} & -1.47777644\times10^{-1} & -2.28815777\times10^{-3} \\
 -1.47777644\times10^{-1} & 2.89511515\times10^{-1} & 1.21113564\times10^{-1} \\
 -2.28815777\times10^{-3} & 1.21113564\times10^{-1} & 8.77574771\times10^{-1}
\end{bmatrix}$ } \\
\hline
\end{tabular}

\vspace{0.5em}
{\footnotesize Mixture weights $(\pi_k)$, mean vectors $(\mu_k)$, and covariance matrices $(\Sigma_k)$ for the 11 cluster Gaussian Mixture Model computed in factor space from LO-CAMS data. Rows list parameters for each class.}
\end{minipage}
\end{center}

\end{document}